\pdfoutput=1
\documentclass[acmsmall,screen]{acmart}

\usepackage{etoolbox}
\newbool{fullversion}
\booltrue{fullversion}

\usepackage{listings}
\usepackage{syntax}
\usepackage{color}
\usepackage{pgfplots}
\usepackage[capitalize]{cleveref}
\usepackage{wrapfig}
\pgfplotsset{compat=1.18}
\definecolor{keywordcolor}{rgb}{0.7, 0.1, 0.1}   \definecolor{tacticcolor}{rgb}{0.0, 0.1, 0.6}    \definecolor{commentcolor}{rgb}{0.4, 0.4, 0.4}   \definecolor{symbolcolor}{rgb}{0.0, 0.1, 0.6}    \definecolor{sortcolor}{rgb}{0.1, 0.5, 0.1}      \definecolor{attributecolor}{rgb}{0.7, 0.1, 0.1} 
\newcommand{\sampcert}{SampCert}
\newcommand{\lean}{Lean}
\newcommand{\mathlib}{Mathlib}
\newcommand{\slang}{\textsc{SLang}}
\newcommand{\AWS}{Amazon Web Services}

\newcommand{\slangM}{\texttt{SLang}}
\newcommand{\probPure}{\texttt{probPure}}
\newcommand{\probBind}{\texttt{probBind}}
\newcommand{\probUniform}{\texttt{probUniform}}
\newcommand{\probUniformByte}{\texttt{probUniformByte}}
\newcommand{\probWhile}{\texttt{probWhile}}
\newcommand{\probUntil}{\texttt{probUntil}}

\newcommand{\probWhileCut}{\texttt{probWhileCut}}

\newcommand{\ADP}{ADP}

\DeclareMathOperator{\lap}{Lap}
\DeclareMathOperator{\geo}{Geo}

\newcommand{\mech}{M}
\newcommand{\dist}{\mathit{PMF}}

\AtBeginDocument{}

\setcopyright{cc}
\setcctype{by}
\acmDOI{10.1145/3729294}
\acmYear{2025}
\acmJournal{PACMPL}
\acmVolume{9}
\acmNumber{PLDI}
\acmArticle{191}
\acmMonth{6}
\received{2024-11-14}
\received[accepted]{2025-03-06}

\begin{document}

\ifbool{fullversion}{
\newcommand{\appref}[2]{\cref{#1}}
  \newcommand{\Appref}[2]{\Cref{#1}}
  \newcommand{\appreflong}[2]{\cref{#1}}
  \newcommand{\apprefshort}[2]{\cref{#1}}
  \newcommand{\Apprefshort}[2]{\Cref{#1}}
}{

\newcommand{\appref}[2]{Appendix {#2} of \cite{sampcert:arxiv}}
  \newcommand{\Appref}[2]{Appendix {#2} of \cite{sampcert:arxiv}}

\newcommand{\appreflong}[2]{Appendix {#2} of the full version of this paper \cite{sampcert:arxiv}}

\newcommand{\apprefshort}[2]{Appendix {#2}~\cite{sampcert:arxiv}}
  \newcommand{\Apprefshort}[2]{Appendix {#2}~\cite{sampcert:arxiv}}
}

\title{Verified Foundations for Differential Privacy}

\author{Markus de Medeiros}
\orcid{0009-0005-3285-5032}
\affiliation{\institution{New York University}
  \city{New York}
  \country{USA}
}
\email{mjd9606@nyu.edu}

\author{Muhammad Naveed}
\orcid{0009-0009-0929-9057}
\affiliation{\institution{Amazon}
  \city{Seattle}
  \country{USA}
}
\email{naveedmd@amazon.com}

\author{Tancrède Lepoint}
\orcid{0000-0003-3796-042X}
\affiliation{\institution{Amazon}
  \city{New York}
  \country{USA}
}
\email{tlepoint@amazon.com}

\author{Temesghen Kahsai}
\orcid{0000-0002-4616-5084}
\affiliation{\institution{Amazon}
  \city{Cupertino}
  \country{USA}
}
\email{teme@amazon.com}

\author{Tristan Ravitch}
\orcid{0009-0005-6967-3945}
\affiliation{\institution{Amazon}
  \city{Denver}
  \country{USA}
}
\email{travitch@amazon.com}

\author{Stefan Zetzsche}
\orcid{0009-0001-1304-0613}
\affiliation{\institution{Amazon}
  \city{London}
  \country{United Kingdom}
}
\email{stefanze@amazon.com}

\author{Anjali Joshi}
\orcid{0009-0004-9398-7242}
\affiliation{\institution{Amazon}
  \city{Boston}
  \country{USA}
}
\email{anjalijs@amazon.com}

\author{Joseph Tassarotti}
\orcid{0000-0001-5692-3347}
\affiliation{\institution{New York University}
  \city{New York}
  \country{USA}
}
\email{jt4767@nyu.edu}

\author{Aws Albarghouthi}
\orcid{0000-0003-4577-175X}
\affiliation{\institution{Amazon}
  \city{Madison}
  \country{USA}
}
\email{awsb@amazon.com}

\author{Jean-Baptiste Tristan}
\orcid{0000-0003-2574-7883}
\affiliation{\institution{Amazon}
  \city{Boston}
  \country{USA}
}
\email{trjohnb@amazon.com}
\authornote{Corresponding author}

\renewcommand{\shortauthors}{de Medeiros et al.}

\begin{abstract}
  Differential privacy (DP) has become the gold standard for privacy-preserving data analysis, but implementing it correctly has proven challenging. Prior work has focused on verifying DP at a high level, assuming either that the foundations are correct or that a perfect source of random noise is available. However, the underlying theory of differential privacy can be very complex and subtle. Flaws in basic mechanisms and random number generation have been a critical source of vulnerabilities in real-world DP systems.

In this paper, we present SampCert, the first comprehensive, mechanized foundation for executable implementations of differential privacy. SampCert is written in Lean with over 12,000 lines of proof. It offers  a generic and extensible notion of DP, a framework for constructing and composing DP mechanisms, and formally verified implementations of Laplace and Gaussian sampling algorithms.
SampCert provides (1) a mechanized foundation for developing the next generation of differentially private algorithms,
and (2) mechanically verified primitives that can be deployed in production systems.
Indeed, SampCert's verified algorithms power the DP offerings of \AWS{}, demonstrating its real-world impact.

SampCert's key innovations include:
(1) A generic DP foundation that can be instantiated for various DP definitions (e.g., pure, concentrated, Rényi DP);
(2) formally verified discrete Laplace and Gaussian sampling algorithms that avoid the pitfalls of floating-point implementations; and
(3)
a simple probability monad and novel proof techniques that streamline the formalization.
To enable proving complex correctness properties of DP and random number generation, SampCert makes heavy use of Lean's extensive \mathlib{} library,
leveraging theorems in Fourier analysis, measure and probability theory, number theory, and topology.

\end{abstract}

\ccsdesc[500]{Security and privacy~Logic and verification}
\ccsdesc[500]{Theory of computation~Probabilistic computation}
\ccsdesc[500]{Theory of computation~Program verification}

\keywords{Differential Privacy, Formal Verification}

\maketitle

\section{Introduction}

Differential privacy (DP)~\cite{DR14} has become the gold standard for privacy-preserving data analysis. It has been widely adopted in industry~\cite{apple,desfontaines2020realworld}, government~\cite{abowd2018us}, and academia as a robust framework for publishing sensitive data while providing strong privacy guarantees. Despite its ubiquity, implementing differential privacy correctly has proven to be a significant challenge. Bugs have been, and continue to be, discovered in the random number generation algorithms~\cite{mironov2012significance} and differentially private mechanisms used in real-world systems~\cite{tang2017privacy,lyu2016understanding,jin2022we}.

Prior work has explored the verification and testing of DP~\cite{near2019duet,10.1145/3589207,zhang2016autopriv,wang2019proving,reed2010distance,gaboardi2013linear,barthe2016proving,10.1145/2976749.2978391,10.1145/3428233,farina2021coupled,albarghouthi2017synthesizing,abuah2021dduo,mcsherry2009privacy,ding2018detecting,bichsel2018dp,gehr2016psi,wang2020check,DBLP:conf/sp/RoyHA21,Sato25}.
However, these works assume either that the foundations of differential privacy are correct or that a perfect source of random noise is available, or alternatively, they focus on proving theoretical results about differentially private algorithms, but do not connect to executable implementations. This approach is unsustainable given the increasing complexity of modern definitions of differential privacy, and there is a history of critical DP vulnerabilities originating from subtle flaws in random number generation~\cite{mironov2012significance}.

In this paper, we present SampCert, the first comprehensive, mechanized foundation for executable differential privacy.
SampCert is written in the Lean v4~\cite{moura2021lean} language and theorem prover and uses more than 12,000 lines of proof.
SampCert offers a generic and extensible notion of differential privacy,
a framework for constructing and composing differentially private mechanisms, and
mechanically verified implementations of the discrete Laplace and Gaussian sampling algorithms.

We see SampCert being used in two ways:
(1) the mechanized DP foundation provides researchers with a powerful starting point for developing and proving the correctness of new privacy mechanisms and definitions; and
(2) the verified primitives, like the random sampling algorithms, can be extracted from Lean and directly deployed to increase assurance of differentially private systems.
Indeed, SampCert is used by the differential privacy offering of \AWS{}.

\subsection{Challenges of Verified DP}

Working with the theory of DP and building practical realizations of it encounters many challenges:

\paragraph{Challenge 1: The Many Faces of DP}
The core challenge in working with differential privacy is deciding on the notion of privacy to use.
In addition to the standard definition of \emph{pure} differential privacy~\cite{DMNS06},
there are a plethora of relaxations of this definition that enable the development of more accurate mechanisms while still providing strong guarantees---\emph{approximate} DP~\cite{approxdp}, \emph{Rényi} DP~\cite{mironov2017renyi}, \emph{zero-concentrated} DP~\cite{bun2016concentrated}, etc.
Thus, proofs of mechanisms, as well as practical implementations, tend to fix a definition of DP and assume basic properties about it, e.g., that it composes additively.

\paragraph{Challenge 2: Developing Correct Mechanisms}
Computing interesting statistics (means, histograms, gradients, etc.) in a differentially private manner requires developing mechanisms that carefully add random noise at certain points in the computation, and proving their correctness with respect to a definition of privacy.
Proving that a mechanism adds the correct type and amount of noise to establish a statistical privacy result turns out to be a difficult task.
Indeed, mistakes have been found in commonly used mechanisms like the Sparse Vector Technique~\cite{lyu2016understanding}.

\paragraph{Challenge 3: Sampling Correctly}
Finally, for a correct implementation of DP, one requires a sampling algorithm that is faithful to the actual mathematical definition of the probability distribution, which is usually the Laplace or Gaussian distribution.
This challenge has been a persistent issue for DP libraries~\cite{mironov2012significance}: approximating real-valued sampling algorithms using floating-point numbers has led to critical breaches of privacy.
To avoid this problem, researchers have proposed theories of discrete DP, which use sampling algorithms that operate over the integers exactly~\cite{canonne2020discrete,balcer2017differential}. \\

\sampcert{} addresses each of these issues in turn: we develop an abstract query language that supports generic DP reasoning, a verified mathematical theory for DP (\cref{section:programmingDP}), and a suite of verified random sampling algorithms which can be efficiently executed (\cref{section:sampling} and \cref{section:extraction}).
\sampcert{} can serve as a foundation to easily develop and prove the correctness of sophisticated DP mechanisms, such as the Sparse Vector Technique.

\subsection{Key Ideas}

\begin{figure}
  \includegraphics[scale=0.4]{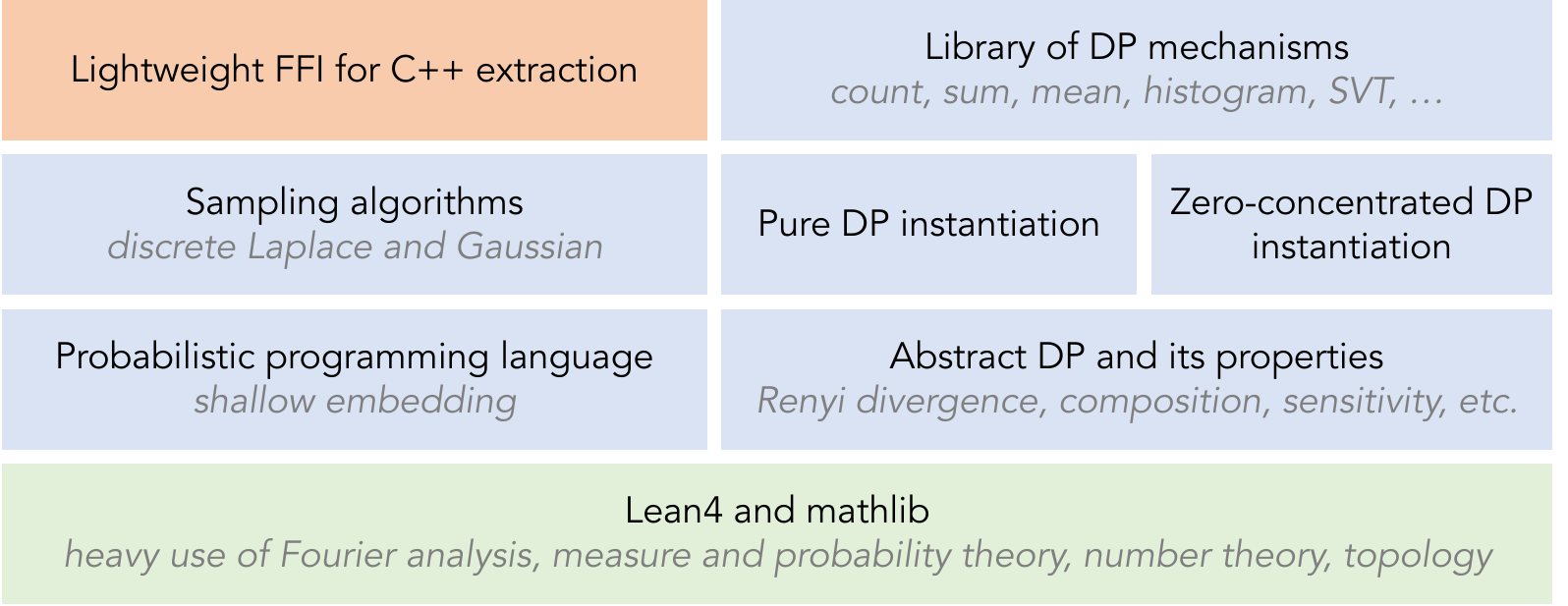}
  \caption{Overview of SampCert's main components, ordered bottom up by dependency.}\label{fig:overview}
\end{figure}

\sampcert{} is designed around several key insights, which put together enable foundational verification of differentially private programs.
\cref{fig:overview} shows the high-level architecture of \sampcert{}.
Below we describe its key ideas.

\paragraph{A Generic DP Foundation}
While the many definitions for DP offer a variety of subtly different statistical guarantees, they also share a set of common properties.
For example, the common definitions of differential privacy include \emph{composition theorems}, which specify upper bounds on the privacy of programs built out of private components.
In \sampcert{}, we collect a suite of these common idioms into an \emph{abstract} interface for DP.
This interface specifies the basic privacy axioms that an instance of DP must satisfy in order to be subject to a generic analysis.
Instead of constraining ourselves to one definition of DP---e.g., pure DP or Rényi  DP---we can develop our mechanisms with respect to the axiomatic definition in a manner parametric to the underlying definition of DP.
In our development, we demonstrate how to instantiate the interface for pure differential privacy and zero-concentrated differential privacy, two of the most widely used definitions of DP.

\paragraph{Random Number Generators (The Heart of DP)}
Differential privacy relies on the ability to sample numbers from, typically, the Laplace or Gaussian distributions. Any mistake in the sampling algorithm may destroy the guarantees of differential privacy.
Unfortunately it is challenging to sample from these distributions correctly. This is particularly true when dealing with floating-point implementations as poignantly demonstrated by~\citet{mironov2012significance}.
Researchers since then have exploited floating-point numbers to mount a series of different attacks~\cite{jin2022we}.

To avoid using floating-point numbers, researchers devised clever algorithms for efficiently sampling from the discrete analogues of the Laplace and Gaussian distributions using exact rational-valued calculations.
In this paper, we present a formally verified implementation of the sampling algorithms presented by \citet{canonne2020discrete}, which are widely used, for instance, by the US Census Bureau for their DP disclosure of statistics.
Specifically, we use Lean to prove that our implementation of the algorithms generates samples following the probability density functions of the discrete Laplace and Gaussian distributions.

\paragraph{A Verification-Ready Probabilistic Programming Language}
A key technical challenge with SampCert is how to cleanly specify the sampling algorithms, DP mechanisms, and theorems, while minimizing the proof burden.
For specifying probabilistic algorithms, we define a simple, functional probabilistic programming language that manipulates mass functions (which need not be probability mass functions). This simple language is represented as a shallowly-embedded monadic DSL in Lean.
By using mass functions rather than full probability spaces (e.g., Hurd's and Giry's monads~\cite{HurdDissertation,frivc2010categorical}), we give ourselves a new proof strategy for analyzing looping programs loosely inspired by \emph{liveness} and \emph{safety} properties.
In particular, our proof strategy allows us to describe the mass distribution of a loop mathematically (as the pointwise limit of the mass functions of its iterates) without requiring that we construct probabilistic loop invariants that may involve complicated normalizing factors.

\paragraph{Extraction and Deployment}
Our simple probabilistic language enables straightforward extraction of Lean code while minimizing the trusted computing base.
Each of the four operators in our probabilistic language maps directly to a simple imperative statement, enabling easy extraction that is sufficiently performant for practical deployment.

One advantage of working with mechanically verified implementations is that we can aggressively optimize our sampling algorithms, without worrying about introducing privacy vulnerabilities.
In SampCert we prove that two Laplace sampling algorithms (originating from two different DP developments) are both correct, allowing us to confidently switch between implementations at runtime.
As we show in \cref{section:performance}, the code we automatically extract from Lean outperforms the implementation of a discrete Gaussian sampler implemented by \citet{canonne2020discrete}.

\paragraph{A Mathematical Buffet}
We make heavy use of Mathlib~\cite{mathlib}, Lean's comprehensive mathematics library, as many proofs require non-trivial mathematics.
By tightly integrating \mathlib{} types into our development, we can leverage this extensive library of mathematical results to reproduce standard differential privacy arguments from the literature, instead of having to develop alternate proofs and techniques that avoid mathematical prerequisites.
For example, since we are dealing with randomized algorithms, we leverage results from \lstinline{Mathlib.Probability} and \lstinline{Mathlib.MeasureTheory}.
Additionally, since we are manipulating non-trivial mathematical expressions, e.g., infinite sums, we make use of \lstinline{Mathlib.NumberTheory} and \lstinline{Mathlib.Topology}.

One concrete example of where this extensive library is essential was in our proof that discrete Gaussian noise establishes zero-concentrated differential privacy (zCDP).
In proving this on paper, \citet{canonne2020discrete} make use of Fourier analysis and the Poisson summation formula, and we are not aware of any simple, alternate proof that avoids this technical machinery.
Fortunately, \mathlib{}'s comprehensive library already includes the relevant theorems from Fourier analysis to replicate this proof.
Using \mathlib{}, we were able to fully formalize this zCDP result from first principles with reasonable proof effort.

\paragraph{Contributions} We summarize our contributions as follows:
\begin{itemize}
\item Mechanically verified foundations of differential privacy in \lean{}, utilizing \mathlib{}.
\item Mechanically verified (discrete) Laplace and Gaussian sampling algorithms.
\item A simple probability monad and novel proof techniques that streamline proof formalization.
\item A Lean-code extractor used to deploy our verified algorithms at \AWS{}.
\end{itemize}

\section{Programming with Differential Privacy}\label{section:programmingDP}

In this section, we describe \sampcert{}'s generic view of differential privacy, how we can verify private mechanisms using our framework, and how we can instantiate it with different DP variants.

\subsection{Abstract Differential Privacy}

Differential privacy is a family of definitions that describe what it means to compute a statistic about a dataset while preserving the privacy of records in that dataset.
For example, suppose that we have the genetic data from a population of consenting individuals and we want to count the number of people with some genetic mutation.
Simply releasing the count is insufficient for protecting the privacy of the individuals.
It has been shown time and time again that typical anonymization techniques are still susceptible to \emph{reconstruction attacks}.
For example, while birth date, gender, or postal code cannot identify an individual alone, their combinations form \emph{quasi-identifiers} which suffice to uniquely identify 87\% of the population of the Unites States~\cite{sweeney2000simple}.

Differential privacy definitions prescribe that adding or removing an individual's genetic data should not change the output statistic significantly.
To achieve this, differential privacy dictates that the output statistics should be randomized.
When implemented correctly, differential privacy ensures that an attacker will require access to an infeasible number of samples in order to learn any information about an individual's presence in the database.

\paragraph{The Mathematics of Privacy}
There are several options for quantifying the privacy risk associated with releasing a randomized statistic.
For our discussion, a mechanism $\mech: T \to \dist(Z)$ is a random function that takes an input in some domain $T$ (typically a database),
and returns a sample from a probability distribution over a countable range $Z$ (with \emph{probability mass function} $\dist(Z)$).
To define privacy, we fix a symmetric ``adjacency'' relation on $T$, and quantify the extent to which releasing a statistic allows an attacker to disambiguate between adjacent values of $T$.
For example when $T$ is the set of databases of genetic information, and databases are considered adjacent when they differ by exactly one row, differential privacy gives an upper bound on the chances that releasing a statistic leaks the genetic data of any individual---even those not included in the dataset.

In its simplest form, \emph{pure} differential privacy says the following:
\begin{definition}[Pure DP]\label{def:puredp}
A mechanism $\mech$ is $\epsilon$-DP if for all adjacent pairs $t,t' \in T$ and $S \subseteq Z$,
\begin{equation}\label{eqn:puredp}
  \Pr[\mech(t) \in S] \leq e^\epsilon \Pr[\mech(t') \in S]
\end{equation}
\end{definition}

\begin{wrapfigure}{r}{0.35\textwidth}
  \vspace{-1em}
  \begin{center}
  \footnotesize
\begin{tikzpicture}
  \begin{axis}[
      width=0.35\textwidth,
      xlabel=$x$,
      ylabel=$f(x)$,
      ymin=0,
      ymax=0.6,
      xmin=-5,
      xmax=5,
      domain=-5:5,
      samples=400,
      legend pos=north west,
  ]
      \addplot[red, thick] {1.0*exp(-abs(x-1))/2};

      \addplot[blue, thick] {1.0*exp(-abs(x))/2};
  \end{axis}
\end{tikzpicture}
\end{center}
  \vspace{-1em}
\caption{Two Laplace distributions.}
\label{fig:dp}
\end{wrapfigure}

In other words, the two probability distributions produced with $t$ or with $t'$ should be close enough, as specified by the \emph{privacy parameter} $\epsilon \in \mathbb{R}$.
The larger $\epsilon$ is, the weaker the privacy guarantee.
This is pictorially shown in \cref{fig:dp}, with two Laplace distributions which have different means.
The closer the distributions are at each point (the smaller the $\epsilon$) the more privacy is provided.
When $\epsilon$ is small, a single random sample does not let an attacker learn much about which of the two distributions it was drawn from, preventing the attacker from learning the true value of our statistic (the mean).

Pure DP is a strong property, because an algorithm must satisfy \cref{eqn:puredp} for all possible configurations $S$ of secret information.
As such, privacy researchers have developed a suite of relaxations to analyze algorithms with less stringent privacy requirements.
One such relaxation is \emph{zero-concentrated} DP (zCDP)~\cite{bun2016concentrated}:

\begin{definition}[zCDP]\label{def:zcdp}
  A mechanism $\mech$ is $\rho$-zCDP if for all adjacent pairs $t,t' \in T$ and $\alpha \in (1,\infty)$,
\[
D_\alpha(\mech(t) || \mech(t')) \leq \alpha\rho
\]
where $D_\alpha$ is the $\alpha$-Rényi divergence between probability distributions.
\end{definition}
In contrast to Pure DP, which gives worst-case upper bounds on the probabilities of identifying events, zero-concentrated DP gives an upper bound on the \textit{expected privacy loss} of a mechanism.
Depending on the application domain, relaxing from the \textit{worst-case} (Pure DP) to an \textit{average case} (zCDP) can be a realistic assumption, making zCDP a popular measure of privacy among researchers and in implementations of DP tools~\cite{tumultanalyticswhitepaper}.
There are several other relaxations based on the expected privacy loss, such as \emph{mean-concentrated} DP~\cite{dwork2016concentrated} and \emph{Rényi} DP~\cite{mironov2017renyi}.

Given the wealth of privacy definitions, it is natural to try and compare their relative strengths.
\emph{Approximate} DP is an even more extreme relaxation of pure differential privacy, which allows a mechanism to violate the privacy specification with probability $\delta$:

\begin{definition}[Approximate DP]\label{def:approxdp}
A mechanism $\mech$ is $(\epsilon,\delta)$-DP if for all adjacent pairs $t,t' \in T$ and all subsets $S \subseteq Z$
we have
\[
\Pr[\mech(t) \in S] \leq e^\epsilon \Pr[\mech(t') \in S] + \delta
\]
\end{definition}

On its own, $(\epsilon, \delta)$-DP does not provide strong privacy guarantees when $\delta$ is non-negligible.
A mechanism may exhibit arbitrarily insecure behavior with probability $\delta$, such as releasing the entire secret database, and still satisfy the definition for $(\epsilon, \delta)$-DP.
While the relative weakness of approximate DP may make it unsuitable for some scenarios, it can nevertheless serve as a basis for comparing the relative strengths of privacy models.
In particular, no useful definition of DP should be \emph{weaker} than approximate DP, so all definitions of privacy should imply an approximate DP bound of some form.

\paragraph{A Unifying Abstraction}
Despite the variety in their definitions, the popular formulations of differential privacy share common characteristics pertaining to the construction and composition of private programs.
Importantly, this means that the correctness of many differentially private algorithms may not depend on the precise variant of DP one may be trying to attain.

Instead of developing separate proof frameworks to verify different privacy bounds, in \sampcert{} we opted for a general, abstract definition of privacy that satisfies a number of basic axioms.
Using our definition, we can build and reason about mechanisms that are parametric in the definition of privacy chosen.
For instance, instead of separately proving privacy for a histogram mechanism for both pure and zCDP, we can construct a single proof using the abstract framework; proofs for pure and zCDP automatically follow, as they are instantiations of the abstract definition.

\begin{figure}
\begin{lstlisting}[caption={Generic DP programs.}, label=lst:dpgeneric, belowskip=-6pt]
abbrev Mechanism (T U : Type) := List T → PMF U
def privComposeAdaptive (nq1 : Mechanism T U) (nq2 : U -> Mechanism T V) (l : List T) :
    PMF (U × V) := do
  let A <- nq1 l
  let B <- nq2 A l
  return (A, B)
def privPostProcess (nq : Mechanism T U) (pp : U → V) (l : List T) : PMF V := do
  let A ← nq l
  return pp A
def privConst (u : U) : Mechanism T U := fun _ => PMF.pure u
\end{lstlisting}
\end{figure}

\begin{figure}
\begin{lstlisting}[caption={Lean definition of AbstractDP. The \lstinline{DiscProbSpace} typeclass aliases \mathlib{} typeclasses for a discrete probability space over a countable, inhabited type. }, label=lst:dpsystem]
class AbstractDP (T : Type) where
  -- A definition of Differential privacy with one monotone real-valued parameter
  -- (eg. γ-DP, γ-zCDP, etc.)
  prop : Mechanism T Z → NNReal → Prop
  prop_mono {m : Mechanism T Z} {γ₁ γ₂: NNReal} :
    γ₁ ≤ γ₂ → prop m γ₁ → prop m γ₂
  -- Privacy bound: sequential composition
  adaptive_compose_prop {U V : Type} [DiscProbSpace U] [DiscProbSpace V]
    {m₁ : Mechanism T U} {m₂ : U → Mechanism T V} {γ₁ γ₂ γ : NNReal} :
    prop m₁ γ₁ → (∀ u, prop (m₂ u) γ₂) → γ₁ + γ₂ = γ →
    prop (privComposeAdaptive m₁ m₂) γ
  -- Privacy bound: pure function postcomposition
  postprocess_prop {U : Type} [DiscProbSpace U]
    { pp : U → V } {m : Mechanism T U} {γ : NNReal} :
    prop m γ → prop (privPostProcess m pp) γ
  -- Privacy bound: constant function
  const_prop {U : Type} [DiscProbSpace U] {u : U} {γ : NNReal} :
    γ = (0 : NNReal) → prop (privConst u) γ
  -- Compatibility: Privacy parameter required to obtain (γ', δ)-approximate DP
  of_app_dp : (δ : NNReal) → (γ' : NNReal) → NNReal
  prop_app_dp [DiscProbSpace Z] {m : Mechanism T Z} : ∀ (δ : NNReal) (_ : 0 < δ) (γ' : NNReal),
    (prop m (of_app_dp δ γ') → ApproximateDP m γ' δ)
\end{lstlisting}
\end{figure}

\cref{lst:dpsystem} depicts the \sampcert{} definition of \emph{abstract differential privacy}, written in terms of the generic DP combinators presented in \cref{lst:dpgeneric}.
From a programming perspective, this typeclass outlines a domain-specific language for constructing and verifying differentially private queries over a list.
Specifically, our abstract definition of differential privacy, which we call $\gamma$-\ADP{}, is parameterized by a real-valued privacy parameter $\gamma$ that takes on a different meaning for each instantiation (e.g., the $\epsilon$ in pure DP or the $\rho$ in zCDP and Rényi  DP).
We define a set of properties that a \lstinline{AbstractDP} instantiation must supply, and informally outline them below:

\begin{enumerate}
\item \textbf{Privacy} (\lstinline{prop}): A specification for what it means for a mechanism $\mech$ to be $\gamma$-\ADP{}, e.g., \cref{def:puredp} for pure DP.
\item \textbf{Monotonicity} (\lstinline{prop_mono}): A proof that if $\mech$ is $\gamma$-\ADP{} then it is $\gamma'$-\ADP{} for all $\gamma' \ge \gamma$.
\item \textbf{Composition} (\lstinline{adaptive_compose_prop}): A proof that \ADP{} mechanisms compose additively.
\item \textbf{Postprocessing} (\lstinline{postprocess_prop}): A proof that postcomposing by functions that do not access the secret database does not degrade privacy.
\item \textbf{Base case} (\lstinline{const_prop}): A proof that constant functions are 0-\ADP{}.
\item \textbf{Approximate DP} (\lstinline{prop_app_dp}): A proof that \ADP{} implies approximate DP (see below).
\end{enumerate}

Most of the above properties, e.g., composition and postprocessing, are standard to definitions of differential privacy.
Perhaps the non-trivial one is the connection with approximate DP.
Here we require that there is a function $f$ (given by the \lstinline{of_app_dp} field) with the property that, for all mechanisms $M$, and parameters $\gamma$ and $\delta$, if $M$ is $f(\gamma, \delta)$-\ADP{}, then $M$ is $(\gamma, \delta)$-DP.

Intuitively, the function $f$ establishes a reduction between an abstract \ADP{} and approximate DP: any $(\gamma, \delta)$-DP privacy requirement can be satisfied by proving some \ADP{} bound.
In the pure DP case, $f(\gamma,\delta) = \gamma$, as any $\gamma$-DP mechanism is $(\gamma,\delta)$-DP for any $\delta$.
For zCDP, we use Lemma 3.5 of \cite{bun2016concentrated} to establish that a $\rho$-zCDP mechanism is $(\rho + \sqrt{4 \rho \log(1/\delta)}, \delta)$-DP for any $\delta$ (indeed, we were able to replicate the proof of this bound from \cite{bun2016concentrated} using \mathlib{} lemmas such as Markov's inequality and the calculus of hyperbolic trigonometry).
While this property of \lstinline{AbstractDP} is not used in our constructions of abstract differential privacy mechanisms, it helps ensure that our instances of \lstinline{AbstractDP} are consistent with respect to each other and standard models of DP.

\paragraph{Notes on the Lean Definitions}
We walk the reader through some of the Lean definitions for clarity.
Consider \lstinline{prop}: it defines a proposition (\lstinline{Prop}) that takes a mechanism (\lstinline{Mechanism T Z}) and a real number $\mathbb{R}_{\geq 0}$ (\lstinline{NNReal} in Lean), and checks if the mechanism satisfies $\gamma$-\ADP{}.
Now consider \lstinline{prop_mono}: it defines a proposition that says if $\gamma_1 \leq \gamma_2$ and the mechanism \lstinline{m} is $\gamma_1$-\ADP{} (denoted \lstinline{prop m }$\gamma_1$), then \lstinline{m} is also $\gamma_2$-\ADP{}.
Consider now \lstinline{adaptive_compose_prop}: it specifies that given two mechanisms that are $\gamma_1$-\ADP{} and $\gamma_2$-\ADP, composing them produces a $(\gamma_1+\gamma_2)$-\ADP{} (the adaptive composition function \lstinline{privComposeAdaptive} is not shown).

\begin{figure}
  \begin{lstlisting}[caption={Lean definition of DPNoise.}, label=lst:dpnoise]
  class DPNoise (dps : AbstractDP T) where
    -- A noise mechanism with sensitivity parameter (Δ) and security parameter (num/den)
    -- (eg. Discrete Laplace, Discrete Gaussian, etc.)
    noise : (query : List T → ℤ) → (Δ  : ℕ+) → (num : ℕ+) → (den : ℕ+) → Mechanism T ℤ
    -- Relationship between noise argument and privacy amount
    noise_priv : (γn : ℕ+) → (γd : ℕ+) → (priv : NNReal) → Prop
    noise_prop {q : List T → ℤ} {Δ γn γd : ℕ+} {γ : NNReal} :
      noise_priv γn γd γ →
      sensitivity q Δ →
      dps.prop (noise q Δ γn γd) γ
\end{lstlisting}
  \vspace{-2pt}
  \end{figure}
\subsection{Noise and Sensitivity}\label{sec:noise}
While \cref{lst:dpsystem} describes how to compose privacy bounds of differentially private programs, it does not directly give us a way to initially obtain privacy bounds for nontrivial programs.
Definitions of DP are typically presented alongside a \emph{noise mechanism}, which prescribes a distribution and amount of noise to add to a statistic in order to obtain a given privacy bound.
The amount of noise depends on both the desired privacy bound and the \textit{sensitivity} of the statistic: statistics whose output changes greatly between adjacent databases will require more noise in order to satisfy a given privacy bound.
In \sampcert{}, we define sensitivity of a function \lstinline{List T → ℤ} to be the maximum change in the value of the function between two adjacent inputs.

\cref{lst:dpnoise} defines a typeclass \lstinline{DPNoise} that specifies the properties of an abstract noising scheme.
The typeclass is parameterized by a definition of differential privacy, and allows provers to construct private noised statistics using the abstract noise mechanism \lstinline{noise}.
The \lstinline{noise} function takes a rational parameter \lstinline{γn/γd}, a sensitivity \lstinline{Δ}, and a \lstinline{Δ}-sensitive query, and adds enough noise to the query to ensure that its result is $\gamma$-\ADP{} (\lstinline{noise_prop}).
We note that a noise program with arguments \lstinline{γn/γd} does not simply provide (\lstinline{γn/γd})-\ADP{} in general (for example, the \emph{Gaussian} noise mechanism with arguments \lstinline{γn/γd} satisfies $(\gamma\textrm{\lstinline{n}}/2\gamma\textrm{\lstinline{d}})^{2}$-zCDP).
The relationship between the desired privacy bound $\gamma$ and the function parameters \lstinline{γn} and \lstinline{γd} is specified in the \lstinline{noise_priv} field.
This setup avoids the use of any floating-point arithmetic, instead requiring that the prover either manually or programmatically ensure that their \lstinline{noise} parameters satisfy the appropriate \lstinline{noise_priv} bound.\footnote{Our most complicated mechanisms in SampCert required at most two proofs of \lstinline{noise_priv}.}

\subsection{Case study: Implementing and Verifying a Private Histogram}\label{sec:histogram}
We now write and verify a simple differentially private program for computing histograms using \sampcert{}.
In particular, we will show that the program is $\gamma$-\ADP{}, meaning that if someone defines a new notion of privacy as an instantiation of $\gamma$-\ADP{}, they get a verified mechanism for free.

A \emph{histogram} over a list is a finite vector of integers (so-called \emph{bins}), which count how elements in a list are assigned to each bin by some \emph{binning function} \lstinline{bin}:

\begin{minipage}{\linewidth}
\begin{lstlisting}
structure Bins (T : Type) (nBins : ℕ) where
  bin : T → Fin nBins
structure Histogram (T : Type) (nBins : ℕ+) (B : Bins T nBins) where
  count : Mathlib.Vector ℤ nBins
\end{lstlisting}
\end{minipage}

Differentially private histograms are a common building block for the construction of larger DP algorithms.
For example, one can privately calculate the approximate maximum of a list by inspecting the last inhabited bin in its histogram, an important step in calculating differentially private means on data whose values lack tight upper bounds a priori~\cite{wilson2019differentially}.
The key idea in constructing a private histogram is to add enough noise to make the value in each bin $(\gamma/\textrm{\lstinline{nBins}})$-\ADP{}; by composition, the overall histogram will be $\gamma$-\ADP{}.

\begin{figure}[t]
\begin{lstlisting}[caption=An implementation of an abstract DP histogram., label=fig:histogram]
variable {T : Type} (nBins : ℕ+) (B : Bins T nBins) [dps : AbstractDP T] [dpn : DPNoise dps]
def privNoisedBinCount (γ₁ γ₂ : ℕ+) (b : Fin nBins) : Mechanism T ℤ :=
  (dpn.noise (exactBinCount nBins B b) 1 γ₁ (γ₂ * nBins))
def privNoisedHistogramAux (γ₁ γ₂ : ℕ+) (n : ℕ) (Hn : n < nBins) : Mechanism T (Histogram T nBins B) :=
  let privNoisedHistogramAux_rec :=
    match n with
    | Nat.zero => privConst (emptyHistogram nBins B)
    | Nat.succ n' => privNoisedHistogramAux γ₁ γ₂ n' (Nat.lt_of_succ_lt Hn)
  privPostProcess
    (privCompose (privNoisedBinCount nBins B γ₁ γ₂ n) privNoisedHistogramAux_rec)
    (fun z => setCount nBins B z.2 n z.1)
def privNoisedHistogram (γ₁ γ₂ : ℕ+) : Mechanism T (Histogram T nBins B) :=
  privNoisedHistogramAux nBins B γ₁ γ₂ (nBins - 1) (proof_pred_nBins_lt_nBins)
\end{lstlisting}
\end{figure}

This argument can be translated one-to-one to a verified implementation of a private histogram in \sampcert{}.
\Cref{fig:histogram} depicts an abbreviated implementation of the abstract histogram function.
Our implementation is parameterized by a positive natural number \lstinline{nBins}, a binning strategy \lstinline{B}, an abstract DP system \lstinline{dps} and abstract noising function \lstinline{dpn}.
Calculation of the overall histogram (in \lstinline{privNoisedHistogramAux}) recursively counts and adds noise to the value for each bin using the function \lstinline{privNoisedBinCount}.
The implementation uses generic DP programs \lstinline{privCompose} to sequence the private operations, and \lstinline{privPostProcess} to update the resulting histogram value with the new count.
This generic construction will enable a generic privacy analysis of \lstinline{privNoisedHistogram}.

To prove that this histogram is $\gamma$-\ADP{}, we begin by showing that calculating the noised count is $(\gamma/\textrm{\lstinline{nBins}})$-\ADP{}.
In \cref{lst:noised} we prove that the exact count in each bin has a sensitivity of 1, so that \lstinline{dpn.noise_prop} can establish the desired DP bound.

\begin{figure}[t]
\begin{lstlisting}[caption=A proof that noised count is $(\gamma/\textrm{\lstinline{nBins}})$-DP., label=lst:noised, belowskip=-10pt]
variable (γ₁ γ₂ : ℕ+) (γ : NNReal) (HN_bin : dpn.noise_priv γ₁ (γ₂ * nBins) (γ / nBins))
theorem exactBinCount_sensitivity (b : Fin nBins) : sensitivity (exactBinCount nBins B b) 1 := by
  rw [sensitivity]
  intros _ _ H
  cases H
  all_goals simp_all [exactBinCount, exactBinCount, List.filter_cons]
  all_goals aesop
lemma privNoisedBinCount_DP  (b : Fin nBins) :
  dps.prop (privNoisedBinCount nBins B γ₁ γ₂ b) (γ / nBins) := by
  unfold privNoisedBinCount
  apply dpn.noise_prop HN_bin
  apply exactBinCount_sensitivity
\end{lstlisting}
\end{figure}

With this result in hand, we prove a privacy bound for \lstinline{privNoisedHistogramAux} by induction.
The base case (an empty histogram) is handled by \lstinline{dps.const_prop}.
For the inductive case, the lemmas \lstinline{dps.postprocess_prop} and \lstinline{dps.compose_prop} break down the proof into establishing a privacy bound of \lstinline{(γ / nBins)} for each bin, which we have done, and \lstinline{(n * γ / nBins)} for the recursive call, which we get from the induction hypothesis as shown in \cref{lst:ind}.

\begin{figure}
\begin{lstlisting}[caption=Proving a privacy bound for \lstinline{privNoisedHistogramAux} by induction., label=lst:ind, belowskip=-10pt]
lemma privNoisedHistogramAux_DP (n : ℕ) (Hn : n < nBins) :
  dps.prop (privNoisedHistogramAux nBins B γ₁ γ₂ n Hn) (n.succ * (γ / nBins)) := by
  induction n
  · unfold privNoisedHistogramAux
    simp
    apply dps.postprocess_prop
    apply dps.compose_prop (AddLeftCancelMonoid.add_zero _)
    · apply privNoisedBinCount_DP; apply HN_bin
    · apply dps.const_prop; rfl
  · rename_i _ IH
    simp [privNoisedHistogramAux]
    apply dps.postprocess_prop
    apply dps.compose_prop ?arithmetic
    · apply privNoisedBinCount_DP; apply HN_bin
    · apply IH
    case arithmetic => simp; ring_nf
\end{lstlisting}
\end{figure}

\begin{figure}
\begin{lstlisting}[caption=Top level privacy bound for \lstinline{privNoisedHistogram_DP}., label=lst:histogramtop]
lemma privNoisedHistogram_DP :
  dps.prop (privNoisedHistogram nBins B γ₁ γ₂) γ := by
  unfold privNoisedHistogram
  apply (AbstractDP_prop_ext _ ?HEq ?Hdp)
  case Hdp => apply privNoisedHistogramAux_DP; apply HN_bin
  case HEq => simp [predBins, mul_div_left_comm]
\end{lstlisting}
\end{figure}

Finally, in \cref{lst:histogramtop} we establish the top-level privacy bound, which follows by simple arithmetic.
This completes the privacy proof for the private histogram, giving us a result that applies to any \lstinline{AbstractDP} instantiation in fewer than 50 lines of proof.
In our development we use this implementation to privately compute approximate maximum and approximate mean queries, their privacy proofs are of a similar level of complexity and reuse the bound proven in \cref{lst:histogramtop}.
This example demonstrates how abstract privacy reasoning is enough to verify the privacy of core DP algorithms, and that we can obtain useful privacy results without fixing a definition of DP beforehand.

\subsection{Instantiating Pure DP}
We now discuss how \lstinline{AbstractDP} can be instantiated into standard definitions of privacy.
To instantiate \lstinline{AbstractDP} to pure differential privacy, we need to establish the properties outlined in \cref{lst:dpsystem}.

\paragraph{Defining Pure DP}
We will give a flavor of some of those definitions and theorems.
First, we define pure differential privacy, which is a Lean formalization of $\epsilon$-DP from \cref{def:puredp}:

\begin{minipage}{\linewidth}
\begin{lstlisting}
def PureDP (m : Mechanism T U) (ε : ℝ) : Prop :=
  ∀ l₁ l₂ : List T, Neighbour l₁ l₂ → ∀ S : Set U,
  (∑' x : U, if x ∈ S then m l₁ x else 0) / (∑' x : U, if x ∈ S then m l₂ x else 0) ≤ ENNReal.ofReal (Real.exp ε)
\end{lstlisting}
\end{minipage}

Using this definition, we can instantiate an \lstinline{AbstractDP} instance for pure DP\footnote{Note that we use a list representation for our database, and have fixed a neighboring relation \lstinline{Neighbour}.} that uses \lstinline{PureDP} for its \lstinline{prop} field.
Our use of the extended nonnegative real numbers \lstinline{ENNReal} means that we do not need to prove convergence of the sums in the above definition---all sums in \lstinline{ENNReal} are absolutely convergent.
The proofs of the \lstinline{AbstractDP} properties (e.g., \lstinline{adaptive_compose_prop}, \lstinline{postprocess_prop}) are standard, and we refer the interested reader to our development.

\paragraph{Laplace Mechanism}
To provide the basic noise mechanism for pure DP, we use discrete Laplace noise, as defined below:

\begin{minipage}{\linewidth}
\begin{lstlisting}
def privNoisedQueryPure (query : List T → ℤ) (Δ : ℕ+) (ε₁ ε₂ : ℕ+) (l : List T) : PMF ℤ := do
  DiscreteLaplaceGenSamplePMF (Δ * ε₂) ε₁ (query l)
\end{lstlisting}
\end{minipage}

This says: apply the $\Delta$-sensitive query to the list \lstinline{l} and add Laplace noise with variance
$\Delta\epsilon_2 / \epsilon_1$.
We can then show that this provides $\epsilon_1/\epsilon_2$-DP, as specified in the following theorem:
\begin{lstlisting}
theorem privNoisedQueryPure_DP (query : List T → ℤ) (Δ ε₁ ε₂ : ℕ+) (bounded_sensitivity : sensitivity query Δ) :
  PureDP (privNoisedQueryPure query Δ ε₁ ε₂) (ε₁ / ε₂)
\end{lstlisting}
(We redact the proof as it involves close to 100 lines of code.)

Using this theorem we are able to establish the abstract properties required by \lstinline{DPNoise}, and with these two instances in hand our generic proofs of privacy (e.g., \cref{sec:histogram}) now establish pure differential privacy when noise is drawn using our discrete Laplace sampler.

\subsection{Instantiating zCDP}
We now discuss how we instantiate \lstinline{AbstractDP} to $\rho$-zCDP (zero-concentrated differential privacy).
This notion, now a standard definition of differential privacy, was presented in \cref{def:zcdp}.

Recall that the definition of zCDP stipulates that
$
D_\alpha(\mech(t) || \mech(t')) \leq \alpha\rho
$,
where $D_\alpha$ is the $\alpha$-Rényi divergence between probability distributions. 
In order to formalize zCDP in Lean, we also formalized the notion of Rényi divergences.
Establishing the properties of $\rho$-zCDP is much more involved and requires definitions of Rényi divergence, integrals, infinite sums, and Jensen's inequality.
Altogether, defining an \lstinline{AbstractDP} instance for zCDP amounts to around 3000 lines of code, however, the proofs themselves closely follow those presented by \citet{bun2016concentrated}.

\paragraph{Gaussian Mechanism}
Analogous to pure DP, we can establish zCDP by adding noise from the discrete Gaussian distribution,
as defined below:
\begin{lstlisting}
def privNoisedQuery (query : List T → ℤ) (Δ : ℕ+) (ρ₁ ρ₂ : ℕ+) (l : List T) : PMF ℤ := DiscreteGaussianGenPMF (Δ * ρ₂) ρ₁ (query l)
\end{lstlisting}
and establish that the result is $\left((1/2) *(\rho_1/\rho_2)^{2}\right)$-zCDP:

\begin{minipage}{\linewidth}
\begin{lstlisting}
theorem privNoisedQuery_zCDP (query : List T → ℤ) (Δ ρ₁ ρ₂ : ℕ+) (bounded_sensitivity : sensitivity query Δ) :
  zCDP (privNoisedQuery query Δ ρ₁ ρ₂) ((1/2) * (ρ₁ / ρ₂)^2)
\end{lstlisting}
\end{minipage}

This theorem allows us to establish a \lstinline{DPNoise} instance for zCDP (where \lstinline{noise ρ₁ ρ₂ ρ} is defined to be \lstinline{(1/2) * (ρ₁ / ρ₂)^2 ≤ ρ}), meaning our generic DP proofs now prove zCDP as well.

\subsection{Additional Differential Privacy Results}

While our \lstinline{AbstractDP} framework is a general and powerful technique for proving DP for compositions of private programs, we remark that not every program in SampCert needs to be proven fully abstractly.
In \appreflong{section:sparsevector}{A} we establish a Pure DP bound on the \emph{sparse vector technique} from~\citet{DR14}, a program which can calculate approximate maximums using asymptotically less noise than private histograms, but to the best of our knowledge is not subject to a generic privacy analysis.
Our development also contains a mechanization of a theorem from~\citet{bun2016concentrated} which establishes zCDP bounds on Pure DP programs, indirectly giving us a zCDP bound on the sparse vector technique as well.
\Apprefshort{section:parallelcomposition}{B} discusses an extension to \lstinline{AbstractDP} for \emph{parallel composition}, and demonstrates how it improves on the noised histogram privacy bound.

Notably, our \lstinline{AbstractDP} framework does not restrict us from using programs that lack a fully abstract proof.
By parameterizing over a generic noised maximum function and its privacy bound, programs that use noised maximums can still be subject to the abstract privacy analysis we presented in \cref{sec:histogram}--only proving the DP-specific privacy bounds for critical components after fixing a definition of privacy.
We believe that the ability to compose with proofs outside of the abstract system will enable SampCert to keep up with the latest developments in DP.

\section{Constructing Verified Sampling Algorithms}\label{section:sampling}

In this section, we discuss our implementation and proof of the \emph{discrete} Laplace and Gaussian algorithms introduced by \citet{canonne2020discrete},
which sample from integer-valued analogues of the real-valued Laplace and Gaussian mechanisms.
By using discrete mechanisms rather than rounded floating-point approximations we can go beyond proofs of their privacy and provide \emph{foundational proofs of their correctness}, describing their posterior distribution exactly.\footnote{To contrast, the clamping mechanism proposed by~\citet{mironov2012significance} privately samples from an approximation of the Laplace distribution, but to the best of our knowledge its exact posterior distribution is not known. }
We hope that this aspect of our development will be useful outside of its applications to verified differential privacy.

Verifying sampling algorithms comes with a different set of design considerations than verifying DP.
In \cref{section:programmingDP} we represented differentially private programs by a shallow embedding into \lean{}'s \lstinline{PMF} (probability mass function) type, and we saw how this allowed us to state and prove DP properties with relative ease.
Unfortunately, \lstinline{PMF} is not a suitable type for verifying sampling algorithms: terms in \lstinline{PMF} describe only the probability mass a program takes at each point, and carry no information about how to sample from the distribution itself.
Indeed, the \lstinline{PMF} type lives within a \lstinline{noncomputable} section in \lean{};
the \lean{} compiler does not come equipped with techniques for compiling a \lstinline{PMF} into executable code.

To address this issue we introduce a new domain-specific programming language \slang{}, which is shallowly embedded in a mass function monad \slangM{}, and which consists solely of terms that \lean{} understands how to compile.
Our architecture balances three needs: (1) minimizing the amount of trusted compilation code required to execute our programs, (2) allowing complete programs to behave like PMF's, and (3) supporting ergonomic reasoning principles during the intermediate stages of a correctness proof (for example, reasoning about the state of a partly unrolled loop).

\subsection{A Simple, Probabilistic Language}
\newcommand{\bind}{\mathbin{>\!\!\!>\mkern-6.7mu=}}

We embed \slang{} in the unnormalized, discrete Giry monad~\cite{frivc2010categorical}.
The general measure-theoretic version of this monad is already formalized in \lstinline{Mathlib.Probability} as part of its existing probability development.
This allows us to apply standard results from probability and measure theory such as Jensen's or Markov's inequality to \slang{} programs, without redeveloping proofs for these results.

Concretely, $\slangM{} \; \tau$ is the type $\tau \to \mathbb{R}_{\ge 0}^{\infty}$; it has monadic return $\delta_{\cdot}$ and bind $(\cdot \bind \cdot)$ defined as
\begin{align}
  \delta_{v'}(v) \triangleq
      \begin{cases}
        1 & v = v' \\
        0 & \textrm{otherwise}
      \end{cases} \\
  (p \bind f)(v) \triangleq \sum_{t \in \tau} \, f(t)(v) \cdot p (t) \label{eqn:bind}
\end{align}

Formally, the series in \cref{eqn:bind} is defined as a sum inside a \mathlib{} topological monoid; the sum always converges absolutely since it takes values in the extended nonnegative reals.
\slangM{} terms with a proof of normalization (i.e., terms with sum 1) can be promoted into \lstinline{PMF} terms, and used to instantiate our abstract DP system.

One may question as to why we express \slang{} programs inside a mass monad with no requirements on the total mass, when we eventually require a proof that a distribution has mass 1.
The principal issue arises when attempting to verify the posterior mass distribution of programs which include loops and recursion.
Consider expressing an invariant on the normalized probability distribution of a program involving loops---for example, a description of the normalized mass distribution at the start of the $k^{\textrm{th}}$ iteration of the loop.
Representing the precise mass function as a \lstinline{PMF} would involve calculating a normalizing factor related to the conditional probability that the loop does not terminate during the first $k-1$ iterates, a problem that involves reasoning quantitatively about how the loop body changes the mass of all possible program states changes rather than just those that are relevant for functional correctness.
Indeed, such a representation requires that all loops be proven to terminate almost-surely before they are fully defined, further complicating the process.\footnote{It would be particularly challenging to apply indirect methods when reasoning about the termination probability, for example by bounding the cumulative density function of the loop's posterior distribution below by a sequence that converges to 1. In our development, we used this technique in our proof of normalization for the sparse vector technique. }
We have found that such complete descriptions of loop invariants to be both uncommon in the literature and counterintuitive to derive ourselves.

Fortunately, this proof burden is almost entirely avoidable when verifying programs whose PMF has a known closed form.
If we are only required to prove that complete \slang{} programs normalize, then we are free to delay their proof of normalization until \emph{after} we have proven their functional correctness.
Ordering our proofs in this way means that the normalization proofs for our random samplers are subject to textbook analysis of well-known mathematical functions: for example, we prove that our Gaussian sampler normalizes by computing the sum of the closed form for its \lstinline{PMF}.
These arguments are straightforward to formalize in \lean{}, using the rich mathematical reasoning principles afforded to us by \mathlib{}.\footnote{Much like normalization, proving convergence for infinite series over arbitrary types is both difficult and unnecessary. By allowing our distributions to take values in $\mathbb{R}^{\infty}_{\ge 0}$ (a type where all series converge), we can delay the proof that they are finite everywhere past the point where we have proven functional correctness. Indeed, because our probability spaces are discrete, the fact that a PMF converges to a finite value at each point follows as a corollary of normalization. }

\begin{figure}[t!]
\footnotesize
\begin{align*}
  \probPure{} & : \tau \to \slangM{}\; \tau \\
  \probBind{} & : (\slangM{}\;\tau) \to (\tau \to \slangM{}\; \tau') \to \slangM{}\; \tau' \\
  \probUniformByte{} & : \slangM{}\; \mathbb{N} \\
  \probWhileCut & : (\tau \to \textrm{\lstinline{Bool}}) \to (\tau \to \slangM{}\; \tau) \to \mathbb{N} \to \slangM{}\;\tau \to  \slangM{}\;\tau \\
  \probWhile & : (\tau \to \textrm{\lstinline{Bool}}) \to (\tau \to \slangM{}\; \tau) \to \slangM{}\;\tau \to  \slangM{}\;\tau \\
  \\
  \probPure{}\; v'\; (v) & \triangleq \delta_{v'}(v) \\
  \probBind{}\; p\; f\; (v) & \triangleq (p \bind f)(v) \\
  \probUniformByte{} (v) & \triangleq
    \begin{cases}
      2^{-8} & v < 2^{8} \\
      0 & \textrm{otherwise}
    \end{cases} \\
  \probWhileCut \; c \; f \; n \; i \; (v)
  & \triangleq
    \begin{cases}
      0 & n = 0 \\
      \probPure{}\; i\; (v) & n > 0 \textrm{ and } \textrm{\lstinline{c v}} = \textrm{\lstinline{false}} \\
      \begin{array}{l}
        \hspace{-5pt} (\probBind\; \\
          \quad (\probBind\; i \; f) \\
          \quad (\probWhileCut\; c\;  f\;  (n - 1))) (v)
        \end{array} & \textrm{otherwise}
    \end{cases} \\
  \probWhile \; c \; f \; i \; (v)
  & \triangleq \sup_{n \in \mathbb{N}} \;\, \probWhileCut\; c\; f\; n\; i \; (v)
\end{align*}
\vspace{-15pt}
\caption{\slang{} operators and their semantics.\vspace{0pt}}\label{fig:slang}
\end{figure}

We can now define our programming language \slang{}, embedded inside \slangM{}.
There are four base terms in \slang{} (see \cref{fig:slang}) which resemble operations in an imperative probabilistic programming language.
The first two constructs $\probPure$ and $\probBind$ mirror returning values and sequencing probabilistic programs, and are described using our mass function monad.
The term $\probUniformByte$ is a uniform distribution over the value of a single byte, which we are able to bootstrap into uniform samples over any finite space.

The final and most complex construct $\probWhile$ is defined to be the supremum over finite truncations of a loop.
In our development we prove that $\probWhileCut$---the subdistribution obtained by truncating $\probWhile$ to a fixed number of loop iterations---has monotone increasing mass at each point.\footnote{This monotonicity property would not hold for all loops if we embedded \slang{} inside \lstinline{PMF}. }
We call the truncation to $k$ iterations a $k$ \emph{cut} of the loop.
By the monotone convergence theorem, $\probWhile$ is the limit distribution of $\probWhileCut$, as would be typical for an operational semantics of an imperative probabilistic programming language.

\subsection{A Proof Technique for Loops}
\label{sec:loop-proofs}
\newcommand{\leantt}{\textrm{\lstinline{true}}}
\newcommand{\leanff}{\textrm{\lstinline{false}}}

Enabled by our use of unnormalized mass functions, our approach to verifying probabilistic loops breaks down proofs for programs involving loops into two lemmas:
\begin{enumerate}
  \item \textbf{Cut reachability}: For each possible output $v$, calculate the mass of $v$ at some cut $k_{v}$.
  \item \textbf{Cut stability}: Prove that the mass at $v$ is stable for cuts greater than $k_{v}$.
\end{enumerate}
In other words, to prove that the posterior distribution of a loop has a particular closed form, we first show how each \emph{individual point} in the sample space reaches its desired mass (reachability), and then show that the mass function is preserved for larger cuts of the loop (stability).
This simplifies the limits we need to compute, and enables us to compute these limits gradually and recursively.
We note that this technique does not work in the normalized setting: the normalized probability mass at an output point $v$ may in fact \emph{not} be stable after a fixed number of loop iterates even if it impossible for the loop to return $v$ after that point: changing the loop's truncation alters the total probability mass, and thus the normalizing factor at each point.

\subsubsection{Sampling from the Geometric Distribution}
\label{sampling-from-the-geometric-distribution}

To illustrate how we reason about programs involving \probWhile{} in \sampcert{} we outline our verification of a sampler for the \textit{geometric distribution}, a simple building block which will be essential in the construction of our more advanced sampling algorithms.
The geometric distribution is a probability distribution on $\mathbb{N}$, parameterized by a real number $t \in [0, 1)$, and given by the PMF
\begin{equation}
  \geo_{t}(z) =
  \begin{cases}
    0 & z = 0 \\
    (1-t)t^{z-1} & z > 0
  \end{cases}
\end{equation}

\begin{figure}
\begin{lstlisting}[caption=An implementation of a geometric sampler in \slang{}., label=fig:geometric-sampler]
  def geoLoopCond (st : Bool × ℕ) : Bool := st.1
  def geoLoopBody (st : Bool × ℕ) : SLang (Bool × ℕ) := do
    let x ← trial
    return (x,st.2 + 1)
  def probGeometricLoop : SLang (Bool × ℕ) :=
    probWhile geoLoopCond (geoLoopBody trial) (true,0)
  def probGeometric : SLang ℕ := do
    let st ← probGeometricLoop
    return st.2
\end{lstlisting}
\end{figure}
Informally, $\geo_{t}(z)$ describes the probability that in a sequence of independent random events which ``succeed'' with probability $t$, the $z^{\textrm{th}}$ event is the first ``failure''.
\Cref{fig:geometric-sampler} depicts a \slang{} implementation of a sampler from the geometric distribution which performs this experiment.
Our implementation is parameterized by a \lstinline{trial} program from which we can draw unlimited independent and identically distributed boolean samples which are \lstinline{true} with probability $t \in [0, 1)$.
The loop state is a value of type \lstinline{Bool × ℕ}: a tuple consisting of the result of the last trial, and the total number of trials attempted so far.
The top-level program \lstinline{probGeometric} performs the sampling experiment in a loop, and reports the total number of trials attempted once a trial fails.

We will show that the mass of \lstinline{probGeometric} at each point $n \in \mathbb{N}$ equals the value $\geo_{t}(n)$.
To begin, a simple argument shows that \lstinline{probGeometricLoop} never returns a state with flag \lstinline{true} (in this case, the loop would have continued executing) so it suffices to determine the value of \lstinline{probGeometricLoop (false, n)}.
Unrolling the definitions, we need to show
for all $t \in [0,1)$ and $n \in \mathbb{N}$,
\begin{equation}
  \geo_{t}(n) = \sqcup_{k \in \mathbb{N}} \; \textrm{\lstinline{probWhileCut geoLoopCond geoLoopBody k (true, 0) (false, n)}}
\end{equation}

For simplicity, let us denote \lstinline{probWhileCut geoLoopCond geoLoopBody} with the function $F$, whose type is $\mathbb{N} \to \textrm{\lstinline{Bool}} \times \mathbb{N} \to \textrm{\lstinline{Bool}}  \times \mathbb{N} \to \mathbb{R}_{\ge 0}^{\infty}$.
The value $F(k, s_{i}, s_{f})$ is the probability mass associated to transitioning from state $s_{i}$ to state $s_{f}$ using at most $k$ (guarded) unrollings of the loop body.
To prove our main result, we will use cut reachability and stability lemmas.

\paragraph{Cut Reachability}
First, we show that 
for each point $n$, truncating the loop to $n+1$ iterates is enough to ensure that the probability mass of sampling $n$ is equal to $\geo_{t}(n)$.

We will set out to find some truncation $k$ such that $F(k, (\leantt, 0), (\leanff, n))$ is known.
The case of $n = 0$ is straightforward, $F(1, (\leantt, 0), (\leanff, 0)) = 0 = \geo_{t}(0)$ since this program will surely increment the counter at least once.
It suffices to show that for all $m$ and $k$ we have $F(m+2, (\leantt, k), (\leanff, m + k + 1)) = \geo_{t}(m+1)$ by specializing $m = n-1$ and $k = 0$.
We can easily prove this by induction on $m$, thus showing that
truncating the loop to $n+1$ iterates is enough to ensure that the probability mass of sampling $n$ is equal to $\geo_{t}(n)$.

\paragraph{Cut Stability}
For the second part of our argument, we will show that adding extra iterates after $n$ does not change the mass at point $n$.
Intuitively, this is due to the fact that exactly one loop iterate can possibly terminate in state $n$.
This can be formalized into the statement that
\begin{equation*}
  \vspace{-0.03in}(n+1+k, (\leantt, 0), (\leanff, n)) = F(n+1, (\leantt, 0), (\leanff, n))
\end{equation*}
and proven using a similar induction and unrolling as in the first part.

\paragraph{Stability + Reachability = Correctness}
Put together, we have shown that for any $n$ the sequence $\{F(i,(\leantt, 0), (\leanff, n))\}_{i \in \mathbb{N}}$ is eventually constant at value $\geo_{t}(n)$.
It is simple to prove that this sequence is monotone increasing, so the supremum of this sequence is equal to its limit, and we conclude that \lstinline{probGeometric n} samples from the geometric distribution with parameter $t$.
We formalize the argument for the normalization of $\geo_{t}$ separately, and because this PMF is equal to \lstinline{probGeometric} at all points, we conclude that our program is indeed normalizing.

\subsubsection{Other \slang{} derived forms}

\sampcert{} includes verified implementations of other basic probabilistic operators which we will make use of in the following sections.
For example, $\probUntil$ implements rejection sampling using a $\probWhile$ loop:
a $\probUntil$ program repeatedly samples from a distribution until it obtains a value that satisfies a boolean predicate.

\begin{minipage}{\linewidth}
\begin{lstlisting}
  def probUntil (body : SLang T) (cond : T → Bool) : SLang T := do
    let v ← body
    probWhile (λ v : T => ¬ cond v) (λ _ : T => body) v
\end{lstlisting}
\end{minipage}

This allows us to bootstrap $\probUniform{}\; n$, a uniform distribution over the integers $[0, n)$, by repeatedly making calls to $\probUniformByte$ in a $\probUntil$ loop.
The correctness proof for $\probUntil$ closely mirrors the proof in \cref{sampling-from-the-geometric-distribution}, however in the \textit{stability} part of the proof the mass has an exponential relationship to the cut value, rather than constant.

We have also formalized samplers for Bernoulli trials---in particular, a sampling algorithm \lstinline{BernoulliExpNegSample} from \citet{canonne2020discrete} for the Bernoulli trial with parameter $e^{-q}$ where $q$ is rational.
We prove these derived forms correct using our \textit{reachability and stability} technique, and they are an integral component for the following sampling algorithms.

\subsection{The Discrete Gaussian and Laplace}

As discussed in \cref{sec:noise}, the noise distributions for Pure DP and zCDP are the \textit{discrete Laplace} and \textit{discrete Gaussian} distributions, respectively.
In this section, we outline our implementations and proofs of those sampling algorithms.
Making use of the large body of mathematical work in \mathlib{}, we are able to adapt mathematical arguments from the literature to prove their correctness.

\subsubsection{The Discrete Laplace}

The Discrete Laplace distribution is a probability space over $\mathbb{Z}$, described by the PMF with a nonnegative scale parameter $t$.
\begin{equation}
  \lap_{t}(z) = \frac{e^{1/t} - 1}{e^{1/t} + 1} \cdot e^{-\vert z \vert / t}
  \label{eqn:laplace}
\end{equation}

The discrete Laplace with \emph{rational} scale parameter $p/q$ can be implemented in \slang{}, and can serve as the \lstinline{noise} program for a definition of pure differential privacy.

\begin{figure}
\begin{lstlisting}[caption=An implementation of the Laplace sampler using sampling loop \lstinline{discretelaplacesampleloop}., label=fig:laplace-sampler]
def DiscreteLaplaceSample (p q : PNat) : SLang ℤ := do
  let r ← probUntil (DiscreteLaplaceSampleLoop p q) (λ x : Bool × Nat => ¬ (x.1 ∧ x.2 = 0))
  return if r.1 then - r.2 else r.2
\end{lstlisting}
\end{figure}

\paragraph{Implementing Laplace in \slang{}}
In \sampcert{}, we implement two different sampling algorithms from the literature, which have different performance characteristics depending on the value of the scale parameter.
To simplify proofs for these samplers, we factor out parts they have in common.

\Cref{fig:laplace-sampler} depicts the top-level sampling procedure for the \slang{} discrete Laplace sampler.
The program executes a \textit{sampling loop} (\lstinline{DiscreteLaplaceSampleLoop}) to obtain a distribution over $\textrm{\lstinline{Bool}}  \times \mathbb{N}$.
Samples from this space are converted into an integer, treating the boolean as the integer's sign, and resampling the value $(\textrm{\lstinline{true}}, 0)$ in order to avoid double-counting the sample at 0.
If probability distribution of the sampling loop program at $(-, n)$ is $(e^{-q/p})^{n} \cdot (1 - e^{-q/p}) \cdot 2^{-1}
$, then the distribution over $\mathbb{Z}$ after this resampling procedure will agree with $\lap_{p/q}$.

\begin{figure}
\begin{lstlisting}[caption={Sampling loops for the discrete Laplace Sampler. The \lstinline{BernoulliExpNegSample p q} function samples from a Bernoulli distribution with bias $\exp(-p/q)$, not shown here.}, label=fig:laplace-sampling-loop-1]
def DiscreteLaplaceSampleLoop (num : PNat) (den : PNat) : SLang (Bool × Nat) := do
  let v ← probGeometric (BernoulliExpNegSample den num)
  let B ← BernoulliSample 1 2 (Nat.le.step Nat.le.refl)
  return (B, v - 1)
def DiscreteLaplaceSampleLoopIn1Aux (t : PNat) : SLang (Nat × Bool) := do
  let U ← UniformSample t
  let D ← BernoulliExpNegSample U t
  return (U,D)
def DiscreteLaplaceSampleLoopIn1 (t : PNat) : SLang Nat := do
  let r1 ← probUntil (DiscreteLaplaceSampleLoopIn1Aux t) (λ x : Nat × Bool => x.2)
  return r1.1
def DiscreteLaplaceSampleLoop' (num : PNat) (den : PNat) : SLang (Bool × Nat) := do
  let U ← DiscreteLaplaceSampleLoopIn1 num
  let v ← probGeometric (BernoulliExpNegSample 1 1)
  let V := v - 1
  let X := U + num * V
  let Y := X / den
  let B ← BernoulliSample 1 2 (Nat.le.step Nat.le.refl)
  return (B,Y)
\end{lstlisting}
\vspace{-4pt}
\end{figure}

We must now implement a sampling loop procedure.
Our first implementation follows the algorithm in Diffprivlib~\cite{diffprivlib}, implemented in \lean{} by \lstinline{DiscreteLaplaceSampleLoop} in \cref{fig:laplace-sampling-loop-1}.
This program obtains the correct sampling loop distribution using a shifted geometric sampler for the natural number part, and an independent coin flip for the sign.
By our analysis of the geometric sampler, this program both meets the correct mass distribution, and normalizes.

Our second implementation is \lstinline{DiscreteLaplaceSampleLoop'} in \cref{fig:laplace-sampling-loop-1}, based on an algorithm given by \citet{canonne2020discrete}.
This implementation uses a different rejection loop in order to separate the integral and fractional parts of the geometric sampling trial, a change which substantially improves the performance of the sampler for large values of $p/q$.
We were able to translate the proof of correctness from \cite{canonne2020discrete}, into \lean{}, making use of some properties about Euclidean division from \mathlib{}.

\paragraph{Establishing Pure DP}
We next prove that discrete Laplace noise forms an instance of the \lstinline{DPNoise} typeclass for pure differential privacy.
In completing these proofs we use \cref{eqn:laplace}.
Once we have this equation characterizing the PMF, our proof of DP does not need to reason explicitly about the computational parts of the algorithm such as loop invariants, and is able to apply standard arguments from the differential privacy literature directly to our \lean{} implementations.
We conclude that the program \lstinline{DiscreteLaplaceSample num den} satisfies $\mathtt{den}/\mathtt{num}$-DP.

\paragraph{Combining the Two Implementations of Laplace}
Indeed, because the sampling loops have equal distributions, their implementation does not matter for our proofs involving \lstinline{DiscreteLaplaceSample}.
\Cref{fig:sampling} (discussed in \cref{section:extraction}) depicts the performance of our Laplace samplers as we increase the scale parameter.
For small values of the parameter it is faster to use the first sampling loop, but for large domains it is faster to use the latter.
Our top-level implementation of discrete Laplace sampling in \sampcert{} gets the best of both worlds by dynamically switching which implementation to use at runtime.
We implemented this switching far into the proof development of SampCert; the fact that all implementations of \lstinline{DiscreteLaplaceSample} are \textit{equal} meant that retrofitting our codebase with this optimization required only superficial changes to our existing privacy proofs, and significantly improved the overall performance of our samplers.

\subsubsection{The Discrete Gaussian}

\begin{figure}
\begin{lstlisting}[caption={\sampcert{} implementation for the discrete Gaussian sampler. }, label=fig:gaussian-sampler-1]
def DiscreteGaussianSampleLoop (num den t : PNat) (mix : ℕ) : SLang (Int × Bool) := do
  let Y ← DiscreteLaplaceSample t 1 mix
  let C ← BernoulliExpNegSample ((|Y| * t * den) - num)^2 (2 * num * t^2 * den)
  return (Y,C)
def DiscreteGaussianSample (num : PNat) (den : PNat) (mix : ℕ) : SLang ℤ := do
  let r ← probUntil (DiscreteGaussianSampleLoop num^2 den^2 (num.val / den + 1) mix) (λ x : Int × Bool => x.2)
  return r.1
\end{lstlisting}
\end{figure}

The canonical \lstinline{noise} function for zCDP is the \textit{discrete gaussian}, a distribution over the integers with PMF $\mathcal{N}_{\mathbb{Z}}(\mu, \sigma^{2})(x) = e^{-(x-\mu)^{2}/2\sigma^{2}} / N_{\mu, \sigma},$ where $N_{\mu, \sigma}$ is a normalizing constant.
Our \slang{} implementation (depicted in \cref{fig:gaussian-sampler-1}) samples from $\mathcal{N}_{\mathbb{Z}}(0, \sigma^{2})$ using a technique presented in \cite{canonne2020discrete}, by repeatedly sampling from a Laplace distribution.
Our formulation allows us to closely mirror the correctness argument from \cite{canonne2020discrete}.
A lemma for $\probUntil$ ensures that the probability of sampling an integer $z$ will be equal to the probability of \lstinline{DiscreteGaussianSampleLoop} sampling $(z, \emph{true})$ divided by the normalizing constant $N$.
Like the proof in \cite{canonne2020discrete}, we do not require explicit reasoning about loop iterates; our lemmas for $\probUntil$ allow us to prove correctness at a higher level.
Adding a fixed value $\mu \in \mathbb{Z}$ to the result of this sampling procedure shifts the mean, yielding a sample from the distribution $\mathcal{N}_{\mathbb{Z}}(\mu, \sigma^{2})$.

\paragraph{Establishing zCDP}
We also prove that the discrete Gaussian function satisfies zero-concentrated differential privacy.
This proof also mimics an argument presented in \cite{canonne2020discrete}.
To establish the privacy bound, we show an upper bound on the Rényi divergence between two samples from discrete normal distributions with different means.
The key step in the argument bounds the normalizing constant for a shifted discrete Gaussian function above by the normalizing constant for the discrete Gaussian with mean zero.
Proving this fact in the manner of \cite{canonne2020discrete} involves computing the Fourier transform of the unnormalized discrete Gaussian function, and applying the Poisson summation formula.
The \mathlib{} library contains an extensive development for verified Fourier analysis, including the Poisson summation formula, which we were able to apply directly.
Altogether, the abstractions we built over \slang{} enabled us to prove the correctness and privacy of a discrete Gaussian sampling algorithm using standard techniques from the privacy literature.

\section{Obtaining Performant Samplers from Lean}\label{section:extraction}

We have deployed \sampcert{}'s verified sampling algorithms at \AWS{}.
While \sampcert{}'s differentially private query implementations are also executable, we have not yet deployed them, in part because integrating these queries directly into existing, large-scale applications and libraries for differential privacy would be more challenging.
In particular, these queries have a more complex interface and interactions with surrounding library code, and the existing implementations at \AWS{} are integrated as part of a highly optimized database.
In contrast, it is easier to deploy routines for generating random samples, since they have a simple interface and are already typically treated as black box library components.
Since researchers have previously identified bugs in sampling algorithms as sources of privacy vulnerabilities, integrating these samplers is a way to provide incremental benefits from verification.

Deployment of our formally verified sampling algorithms requires providing an interface for projects outside of \lean{} to call and execute our \slang{} samplers.
In \sampcert{} we leverage a small trusted codebase in order to compile \slang{} programs to conventional languages like C++ and Python, which are languages suitable for deployment at scale by \AWS{}.

\subsection{Compilation and Deployment}

Our simple probabilistic language, \slang{}, consists of a small number of monadic operators with direct correspondence to imperative operations.
We have used two approaches for obtaining executable code from \slang{} programs.
The first involves compilation using the \lean{} \emph{foreign function interface} (FFI), allowing us to compile and execute our DP mechanisms while adding only 57 lines of C++ to our trusted computing base.
The second involves a translation from \slang{} to Dafny~\cite{Dafny} and then from Dafny to Python, which we used in order to deploy our verified sampling algorithms at \AWS{}.

\paragraph{Computable Functions}
The \lean{} language makes a clear distinction between \textit{computable} and \textit{noncomputable} definitions.
By default, any \lean{} term whose definition depends on non-computational constructs (e.g., the law of the excluded middle or classical choice) must be marked as \lstinline{noncomputable}, and will be rejected by \lean{}'s C++ compilation pipeline.
The four primitive \slang{} constructs fall into this category, as all \slangM{} terms are functions into the classical real numbers.

Even though \lean{} has no default compilation strategy for \textit{noncomputable} terms, it allows programmers to nevertheless compile programs that use such definitions by using the foreign function interface.
Using an \lstinline{external} annotation, a programmer can specify that any term be compiled to an arbitrary C++ function, enabling the \lean{} compiler to treat it as if it were \textit{computable}.
While \lean{} guarantees that \lstinline{external} definitions will not affect the soundness of the logic (and hence, our library of verified DP results), it can make no formal guarantees about the runtime behavior of the compiled binary.
Therefore, it is paramount that external definitions be kept as minimal as possible.
Our FFI footprint for \sampcert{} consists of one external definition for each \slang{} primitive, and one function to trigger top-level evaluation.
In total, five functions (57 lines of C++) are enough to compile and run \slang{} programs from \lean{}.
Those functions, which correspond directly to \slang{} operators, are partially shown in \cref{lst:cpp} (see \appreflong{section:moreperformance}{C} for a complete listing).
The external function for $\probPure\;v$ simply returns $v$.
The external function for $\probBind\;p\;f$ obtains a value from $p$, and returns the value from executing $f\;p$.
The external function for $\probWhile$ executes a C++ while loop.

Random sampling occurs in the external function for $\probUniformByte$, which reads and returns a single byte from \lstinline{/dev/urandom}.
Reading random bytes, rather than uniform random integers, simplifies the trusted FFI implementation and means that we do not need to perform any subtle, bug-prone bit manipulation in C++.
Instead, in SampCert, we implement and verify the code which simulates uniform samples using a stream of uniformly random bytes in \lean{}.

\begin{figure}[t]
  \small
\begin{lstlisting}[caption={Some C++ external functions for \slang{} operators.}, label=lst:cpp]
extern "C" lean_object * prob_Pure(lean_object * a, lean_object * eta) {
    lean_dec(eta);
    return a;}
extern "C" lean_object * prob_UniformByte (lean_object * eta) {
    lean_dec(eta);
    unsigned char r;
    read(urandom, &r,1);
    return lean_box((size_t) r);}
extern "C" lean_object * prob_Bind(lean_object * f, lean_object * g, lean_object * eta) {
    lean_dec(eta);
    lean_object * exf = lean_apply_1(f,lean_box(0));
    lean_object * pa = lean_apply_2(g,exf,lean_box(0));
    return pa;} 
\end{lstlisting}
\end{figure}

\subsection{Optimizations and Performance Comparison}\label{section:performance}

In this section, we compare the performance of our verified discrete sampling algorithms to two other implementations of samplers for discrete Gaussians:
the discrete Gaussian (\texttt{sample\_dgauss}) implementation supplied by \citet{canonne2020discrete} and another implementation of the same algorithm in \texttt{diffprivlib}~\cite{diffprivlib}.
Because these two implementations are written in Python, for these benchmarks, we use the Python-extracted version of our sampling algorithms (\apprefshort{section:moreperformance}{C} contains more details about this extraction set-up).

\begin{wrapfigure}{r}{.5\textwidth}
  \centering
\includegraphics[scale=0.38]{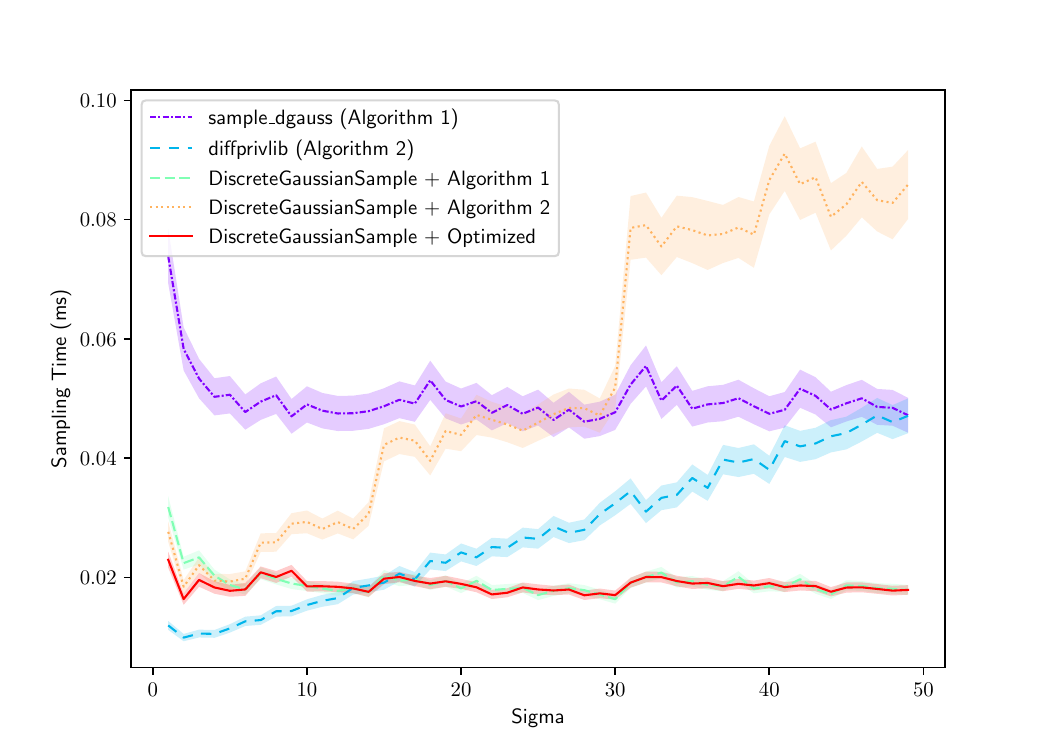}
  \caption{Runtime of \sampcert{}'s Gaussian sampler vs other tools.
    The three runs of \lstinline{DiscreteGaussianSample} are configured to use the Laplace sampling algorithm from \lstinline{sample_dgauss} (Algorithm 1), diffprivlib (Algorithm 2), or switch between them dynamically (Optimized).}\label{fig:sampling}
\end{wrapfigure}

\cref{fig:sampling} shows the runtime (in ms) for the different sampling algorithms as we vary the standard deviation of the Gaussian distribution.

For \sampcert{} we include three variants of the Gaussian samplers:
(green) suited for higher values of the standard deviation, (yellow) suited for lower values of standard deviation, and (red) which dynamically changes its Laplace sampling algorithm based on the standard deviation, to minimize runtime.
All of these are proven correct in Lean and shown in \cref{fig:laplace-sampling-loop-1}.

As shown in \cref{fig:sampling}, the extracted Lean sampler outperforms the implementation of \citet{canonne2020discrete} on this range of parameters, with more than a 2x difference in runtime.
The results also show how our optimized sampler is better than either sampler individually.
The sampler in \texttt{diffprivlib} has a runtime that is linear in the given standard deviation (blue), while our optimized sampler's runtime remains constant.
The \texttt{diffprivlib} sampler is faster than ours for smaller values of the standard deviation.
However, it uses floating-point operations for calculating some of the parameters and constants used in sampling, whereas, SampCert uses rational arithmetic throughout to avoid round-off error.
The plot for our implementation of Algorithm 2 has sharp jumps at successive powers of two of the sigma parameter, which are not seen in the \texttt{diffprivlib} implementation of this algorithm. We believe that these steps are caused by our exact uniform sampling process, which we elaborate on in \apprefshort{section:moreperformance}{C}.
Since Algorithm 1 has better performance for larger values of the sigma parameter, we did not further optimize this aspect of our implementation of Algorithm 2.

\section{Related Work}
We now contrast our work with techniques for verifying and testing differential privacy.

\paragraph{Verifying DP}
There is a large body of work on verifying and enforcing differential privacy.
For much of this work, the primary differences with SampCert are the following:
\begin{enumerate}
\item Existing techniques are typically restricted to a specific notion of privacy, e.g., pure differential privacy, whereas SampCert aims to be generic and extensible.
It is important to note however the linearly typed programming languages Duet~\cite{near2019duet} and Jazz~\cite{10.1145/3589207}, enable reasoning about a family of privacy definitions, including zero-concentrated DP.
\item Existing techniques tend to be constrained by a type system~\cite{zhang2016autopriv,wang2019proving,reed2010distance,gaboardi2013linear}, program logic~\cite{barthe2016proving,10.1145/2976749.2978391,10.1145/3428233}, proof rules~\cite{albarghouthi2017synthesizing}, or runtime system~\cite{abuah2021dduo,mcsherry2009privacy}. These systems tend to restrict the generality of the mechanisms they can verify, in exchange for either more automation or specialized reasoning principles that simplify certain proofs. SampCert is built in Lean and admits arbitrary proofs written in Lean, extending the scope of systems that can be verified in principle, but requires manual proofs. Leveraging mathlib's extensive library helps offset this proof burden.
\item Existing techniques assume correctness of the foundations of differential privacy (e.g., composition theorems), and are taken as axioms of the type system or logic. SampCert, on the other hand, is meant as a verified foundation of DP, and therefore proves properties of DP like adaptive composition from first principles.
\item Finally, all existing techniques assume the existence of a perfect sampling algorithms. SampCert proves correctness of the sampling procedures.
\end{enumerate}  

Some other works have sought to build foundationally verified libraries for differential privacy in theorem provers, similarly to \sampcert{}, that address some of these above limitations.
For example, seeking to address a similar concern over round-off errors in trusted samplers, \citet{amorim-finite-dp} verify the Geometric Truncated Mechanism~\cite{balcer2017differential} which only requires finite precision to sample. 
In contrast to SampCert, they do not derive an extractable implementation based on their verification.
CertiPriv \cite{Barthe12} certifies the differential privacy of a discrete form of the Gaussian mechanism in Rocq, but their proof rule has a side condition that requires showing a challenging Gaussian tail bound. This kind of tail bound is similar to one that \citet{canonne2020discrete} (and our mechanization) needed to use the Poisson summation formula to show.
Unlike \sampcert{}, the CertiPriv development includes no closed examples using this proof rule.

Recently, \citet{Sato25} have developed a foundationally verified library for differential privacy with continuous distributions in Isabelle/HOL.
Just as \sampcert{} makes use of mathlib's extensive library of results, \citet{Sato25} are able to rely on Isabelle/HOL libraries for theorems from analysis and probability theory.
In contrast to \sampcert{}, their work does not develop executable implementations of these algorithms.
They leave the question of how to formally bound the error of floating point approximations of continuous distributions to future work.

\paragraph{Automated Testing of DP}
There is a body of literature focused on discovering bugs in differentially private mechanisms, instead of proofs of correctness. Various search techniques are employed, including optimization and symbolic methods. For instance, StatDP~\citep{ding2018detecting} uses statistical tests for counterexample generation. DP-Finder~\citep{bichsel2018dp} transforms the search into an optimization problem using surrogate functions and numerical methods, with verification done by exact solvers like PSI~\citep{gehr2016psi} or sampling-based estimators. \textsc{CheckDP}~\citep{wang2020check} combines verification and falsification using symbolic methods.
Kolahal~\citep{DBLP:conf/sp/RoyHA21} uses the testing techniques to discover correct noise parameters for privacy mechanisms.

\paragraph{Reasoning about Probabilistic Loops in Shallow Embeddings} A number of prior works reason about randomized programs in theorem provers using a shallow embedding in which randomized programs are written monadically.
In HOL, \citet{HurdDissertation} used a state monad in which randomized programs accessed an infinite tape of uniformly sampled bits.
Given such a program, a probability distribution is obtained by integrating over the space of possible tapes.
\citet{HurdDissertation} defines approximations of while loops that stop looping after a bounded number of iterations, analogous to our cuts, and derives a number of proof rules for deriving behaviors of unbounded loops in terms of the behavior of cuts.
However, as each finite cut is still a normalized probability distribution, the stability and reachability proof technique described in \cref{sec:loop-proofs} is not applicable.
\citet{DBLP:conf/mpc/AudebaudP06} used a monad of \emph{(sub)-measure transformers} in Coq.
To model the semantics of randomized recursion, they axiomatize an $\omega$-CPO structure on the interval $[0, 1]$, and derive induction principles for reasoning about fixed points.
\citet{DBLP:conf/esop/EberlHN15} give a foundational definition of the Giry monad in Isabelle, which can represent general measure-theoretic programs.
Other works use only a finite probability distribution monad~\citep{DBLP:conf/types/WeegenM08}.
This avoids the complexities of reasoning about countable series or measure-theoretic integration, but restricts the kinds of looping constructs that can be modeled, since general probabilistic loops can generate countable distributions.

\section{Limitations and Future Work}

The design decisions behind \sampcert{} introduce some limitations, some which could be addressed in future work:

\paragraph{Countable Domains} All mechanisms and mass functions in \sampcert{} take place in discrete probability spaces over countable types.
While we believe that our privacy development can be translated into a more general measure-theoretic setting (several of our results specialize measure-theoretic results from \mathlib{}), generalizing our proof technique for loops may be more difficult as it relies on characterizing a mass function pointwise.

\paragraph{Support for Floating-Point Algorithms} Floating-point algorithms can in principle be represented using \slang{} programs since the set of floating point numbers is countable. However, since library support for reasoning about floating-point numbers in \lean{} is currently limited and our proofs typically rely on closed-form expressions for PMF's (which are typically unknown for floating-point algorithms), obtaining foundational privacy guarantees about floating-point samplers in \sampcert{} would be a significant challenge.

\paragraph{Database Types} Our DP queries, such as the private histogram, represent a database as plain lists, instead of a more realistic type.
This means that \sampcert{} cannot express some properties such as group privacy. We are interested in exploring this more as future work.

\paragraph{Defining AbstractDP using Approximate DP} We require that all instances of AbstractDP are stronger than approximate DP so that we make privacy claims about all AbstractDP instances at \AWS{}.
Hence we believe that this requirement is safe to remove from AbstractDP if one seeks to verify variants of DP that are weaker than or incomparable to approximate DP.

\paragraph{Single-Parameter DP definitions} Abstract DP only supports variants of differential privacy with one real parameter, as we found that parameterizing it to support multi-parameter DP definitions led to a less usable proof interface.

\section{Conclusion}
We presented \sampcert{}, a formally verified foundation for differential privacy.
SampCert's key innovations include:
(1) A generic DP foundation that can be instantiated for various DP definitions (e.g., pure, concentrated, Rényi DP);
(2) formally verified discrete Laplace and Gaussian sampling algorithms that avoid the pitfalls of floating-point implementations; and
(3)
a simple probability monad and novel proof technique that streamline the formalization.
We see \sampcert{} being used in two ways:
(1) The mechanized DP foundation provides researchers with a powerful starting point for developing and proving correctness of new privacy mechanisms and definitions.
(2) The mechanically verified primitives, like the random sampling algorithms, can be extracted from Lean and directly deployed to increase assurance of differentially private systems.
In the future, we would like to extend \sampcert{} to model and prove non-existence of timing side-channels.
Additionally, we would like to use \sampcert{} as to verify and extract a full-fledged DP query evaluation engine.

\section*{Data Availability Statement}

The reviewed artifact for our paper is available online \cite{artifactv1}, the latest version can be found at \cite{artifact}. The SampCert development is available on GitHub at \href{https://github.com/leanprover/SampCert}{https://github.com/leanprover/SampCert}, on the \lstinline{dev} branch.

\begin{acks}
  The authors thank the anonymous reviewers for their helpful feedback.
  Tassarotti and de Medeiros's contributions to this material are based in part upon work supported by the \grantsponsor{NSF}{National Science Foundation}{} under Grant No.~\grantnum{NSF}{2338317} and an Amazon Research Award.
\end{acks}

\bibliographystyle{ACM-Reference-Format}
\bibliography{references}

\pagebreak

\appendix

\section{The Sparse Vector Technique}\label{section:sparsevector}

While \lstinline{AbstractDP} provides an interface for composing differentially private queries, not all DP constructions are subject to a generic privacy analysis.
One such example is the \textit{Sparse Vector Technique}.
In \sampcert{} we mechanize the proof of pure DP for a non-adaptive version of the sparse vector technique from \citet{DR14}, and establish zCDP using a conversion bound from \citet{bun2016concentrated}.

\subsection{Establishing Pure Differential Privacy}

\newcommand{\clippedsum}{\textrm{\lstinline{clipped_sum}}}

The sparse vector technique can be applied in a scenario where we have a collection $Q$ of sensitivity-1 queries, and we want to find which queries in this collection return a value that exceeds some threshold $T$ when run on a private database $D$.
If $Q$ is a fixed and finite set, then one approach would be to compute $q(D)$ for each $q \in Q$, add noise, and then report whether the result exceeded $T$.
However, if we naively try to apply the composition theorems we have seen so far, the privacy cost would be in proportion to the total number of queries in $Q$.

Instead, the sparse vector technique allows us to incur a privacy cost that is proportional only to the number of queries that actually exceed the threshold.
Moreover, it works even when the queries to be run are not a fixed set $Q$ that is known ahead of time, but are instead an \emph{adaptively} chosen stream of queries, $qs_1, qs_2, \dots$, where later queries in the stream can be chosen based on information about whether earlier queries have exceeded the threshold or not.
For simplicity, to avoid having to model interactivity, we consider here a version that is non-adaptive.
However, we still model that the collection of queries is an \emph{a priori} unbounded stream, as represented by a function \lstinline{qs : ℕ -> Query sv_T ℤ}, where \lstinline{(qs i)} corresponds to the $i$th element of the stream, and \lstinline{sv_T} is the type of the data records in the private database.\footnote{A finite stream of queries of length $n$ can be mimicked by setting  \lstinline{qs i} for \lstinline{i} greater than $n$ to be a constant-valued query that always returns a value far above the threshold $T$.}

\begin{figure}
\begin{lstlisting}[caption={\sampcert{} implementation for the sparse vector technique. }, label=fig:sparsevector-sv1]
def privNoiseZero (ε₁ ε₂ : ℕ+) : SPMF ℤ := dpn.noise (fun _ => 0) 1 ε₁ ε₂ []

def privNoiseGuess (ε₁ ε₂ : ℕ+) : SPMF ℤ := privNoiseZero ε₁ (2 * sens_cov_vk * ε₂)

def privNoiseThresh (ε₁ ε₂ : ℕ+) : SPMF ℤ := privNoiseZero ε₁ (2 * sens_cov_τ * ε₂)

def sv_query (sv_T : Type) : Type := ℕ -> Query sv_T ℤ

def sv1_state : Type := List ℤ × ℤ

def sv1_threshold (s : sv1_state) : ℕ := List.length s.1

def sv1_noise (s : sv1_state) : ℤ := s.2

def sv1_aboveThreshC (qs : sv_query sv_T) (T : ℤ) (τ : ℤ) (l : List sv_T) (s : sv1_state) : Bool :=
  decide (qs (sv1_threshold s) l + (sv1_noise s) < τ + T)
  -- decide (exactDiffSum (sv1_threshold s) l + (sv1_noise s) < τ)

def sv1_aboveThreshF (ε₁ ε₂ : ℕ+) (s : sv1_state) : SLang sv1_state := do
  let vn <- privNoiseGuess ε₁ ε₂
  return (s.1 ++ [s.2], vn)

def sv1_aboveThresh {sv_T : Type} (qs : sv_query sv_T) (T : ℤ) (ε₁ ε₂ : ℕ+) (l : List sv_T) : SLang ℕ := do
  let τ <- privNoiseThresh ε₁ ε₂
  let v0 <- privNoiseGuess ε₁ ε₂
  let sk <- probWhile (sv1_aboveThreshC qs T τ l) (sv1_aboveThreshF ε₁ ε₂) ([], v0)
  return (sv1_threshold sk)
\end{lstlisting}
\end{figure}

\paragraph{Above Threshold Algorithm}
The core building block of the sparse vector technique is the \emph{above threshold} algorithm that returns the first query in the stream that exceeds the threshold (up to the addition of noise).
\Cref{fig:sparsevector-sv1} gives a \sampcert{} implementation of this algorithm, based on Algorithm 1 from \citet{DR14}.
The algorithm first samples noise \lstinline{τ} to add to the threshold \lstinline{T}.
Then, it iteratively runs each query \lstinline{qs i}, adds additional noise to the query's output, and compares whether the result is greater than or equal to \lstinline{τ + T}.
It stops and returns the index of the first query that exceeds this threshold.

The proof of differential privacy in \citet{DR14} proceeds by defining a sequence of random variables $g_{k}(L)$ denoting the maximum noised value of the first $k$ queries.\footnote{\citet{DR14} leave the index on $g_{k}(L)$ implicit but we include it here for clarity.}
To replicate their proof, we progressively manipulate the code in \cref{fig:sparsevector-sv1} into a form which makes $g_{k}(L)$ explicit, culminating in a version called \lstinline{sv9_aboveThreshold}.
To establish the equivalence of \lstinline{sv1_aboveThreshold} and \lstinline{sv9_aboveThreshold}, we establish a chain of equivalences through seven other intermediate versions called \lstinline{sv2_aboveThreshold} through \lstinline{sv8_aboveThreshold}.
We briefly outline a subset of the interesting points in the sequence of manipulations here, and we refer the interested reader to our development for the full implementations and equality proof for each step.

\begin{itemize}
  \item{\lstinline{sv3_aboveThresh}: Obtains the value of the mass function at $n$ using a $\probWhileCut$ loop truncated to $n+1$ iterations. While this $\slangM$ term is \lstinline{noncomputable} (it is not composed of primitive $\slang{}$ terms), it is equal to \lstinline{sv1_aboveThresh} at each point, so its differential privacy implies the privacy of \lstinline{sv1_aboveThresh}. The equality proof involves a cut reachability and stability argument. }
  \item{\lstinline{sv4_aboveThresh}: Commutes the $n+1$ random samples used to calculate the value of the mass function at any point $n$ out of the truncated loop, so that the loop becomes deterministic. }
  \item{\lstinline{sv6_aboveThresh}: Rewrites the loop into a closed form. }
  \item{\lstinline{sv9_aboveThresh}: Isolate the $n=0$ case, so that the $n>0$ cases can be written in terms of $g_{k}$ from \cite{DR14}. }
\end{itemize}

\begin{figure}
\begin{lstlisting}[caption={Sparse vector technique in a form that is subject to privacy analysis.}, label=fig:sparsevector-sv9]
def sv4_presample (ε₁ ε₂ : ℕ+) (n : ℕ) : SLang { l : List ℤ // List.length l = n } :=
  match n with
  | Nat.zero => return ⟨ [], by simp ⟩
  | Nat.succ n' => do
    let vk1 <- privNoiseGuess ε₁ ε₂
    let vks  <- sv4_presample ε₁ ε₂ n'
    return ⟨ vks ++ [vk1], by simp ⟩


def sv8_sum (qs :  sv_query sv_T) (l : List sv_T) (past : List ℤ) (pres : ℤ) : ℤ :=
  qs (List.length past) l + pres

def sv8_G (qs :  sv_query sv_T) (l : List sv_T) (past : List ℤ) (pres : ℤ) (future : List ℤ) : ℤ :=
  match future with
  | []        => sv8_sum qs l past pres
  | (f :: ff) => max (sv8_sum qs l past pres) (sv8_G qs l (past ++ [pres]) f ff)

 def sv8_cond (qs :  sv_query sv_T) (T : ℤ) (τ : ℤ) (l : List sv_T) (past : List ℤ) (pres : ℤ) (future : List ℤ) (last : ℤ) : Bool :=
   (sv8_G qs l past pres future < τ + T) ∧ (sv8_sum qs l (past ++ [pres] ++ future) last ≥ τ + T)

def sv9_aboveThresh (qs :  sv_query sv_T) (T : ℤ) (ε₁ ε₂ : ℕ+) (l : List sv_T) : SLang ℕ :=
  fun (point : ℕ) =>
  let computation : SLang ℕ := do
    match point with
    | 0 => do
      let τ <- privNoiseThresh ε₁ ε₂
      let v0 <- privNoiseGuess ε₁ ε₂
      if (sv8_sum qs l [] v0 ≥ τ + T)
        then probPure point
        else probZero
    | (Nat.succ point') => do
      let v0 <- privNoiseGuess ε₁ ε₂
      let presamples <- sv4_presample ε₁ ε₂ point'
      let τ <- privNoiseThresh ε₁ ε₂
      let vk <- privNoiseGuess ε₁ ε₂
      if (sv8_cond qs T τ l [] v0 presamples vk)
        then probPure point
        else probZero
  computation point
\end{lstlisting}
\end{figure}

\Cref{fig:sparsevector-sv9} shows this final rewritten version of the algorithm.
While this version is now both \lstinline{noncomputable} and seemingly more complicated than \lstinline{sv1_aboveThresh}, it is now expressed in terms of random variables that mirror the proof in \cite{DR14} closely.
Since the combined equality proofs establish that \lstinline{sv9_aboveThesh = sv1_aboveThresh}, it suffices now to prove $\epsilon$-DP of \lstinline{sv9_aboveThresh}.

The proof proceeds as follows.
We state as an assumption that each of the queries in the stream have sensitivity 1,
 by adding a Lean varaible declaration of the form \lstinline{variable (Hqs_sens : ∀ i, sensitivity (qs i) 1)}.
Next, let $D$ and $D'$ be neighboring databases, it suffices to show that for all $n \in \mathbb{N}$,
\begin{equation}
\frac{\textrm{\lstinline{sv9_aboveThresh}}\; qs \; T\; \epsilon_1 \; \epsilon_2 \; D\; n}{\textrm{\lstinline{sv9_aboveThresh}}\; qs \; T\; \epsilon_1 \; \epsilon_2 \; D'\; n} \leq e^{\epsilon_{1}/\epsilon_{2}}.
\end{equation}
We will focus on the $n > 0$ case since the $n=0$ case is a degenerate version of it.
Unfolding the definition of \lstinline{sv9_aboveThresh}, we can factor the calculation of the variable \lstinline{presamples} out of each sum and cancel them out in the quotient--effectively treating all but the calculation of $\tau$ and $vk$ as deterministic.
This leaves us to show that for all choices of \lstinline{presamples} and $v_{0}$,
\begin{equation}
  \frac{\sum_{\tau \in \mathbb{Z}} \sum_{v_{k} \in \mathbb{Z}} \lap_{\epsilon_{1}/2\epsilon_{2}}(\tau) \cdot \lap_{\epsilon_{1}/4\epsilon_{2}}(v_{k}) \cdot [\textrm{\lstinline{sv8_cond}}\; qs \; T\; \tau\; D\; \textrm{\lstinline{List.nil}} \; v_{0}\; \textrm{\lstinline{presamples}}\; v_{k}]} {\sum_{\tau \in \mathbb{Z}} \sum_{v_{k} \in \mathbb{Z}} \lap_{\epsilon_{1}/2\epsilon_{2}}(\tau) \cdot \lap_{\epsilon_{1}/4\epsilon_{2}}(v_{k}) \cdot [\textrm{\lstinline{sv8_cond}}\; qs \; T\; \tau\; D'\; \textrm{\lstinline{List.nil}} \; v_{0}\; \textrm{\lstinline{presamples}}\; v_{k}]} \leq e^{\epsilon_{1}/\epsilon_{2}}
\end{equation}
where the Iverson bracket $[ \cdot ]$ is 1 if and only if its argument is $\leantt$.
This inequality now follows from a change of variables, exactly as in \cite{DR14}, establishing $\epsilon$-DP.
We highlight that aside from the translation from \lstinline{sv1_aboveThresh} to \lstinline{sv9_aboveThresh}, the proof of privacy is entirely standard. \\

\paragraph{Almost-Sure Termination.}
There is one more issue to address: in \sampcert{} we have chosen to only define $\epsilon$-DP for programs which terminate almost surely, and so we must prove that the mass of \lstinline{sv1_aboveThresh} is one.
Since the stream of queries is unbounded, without assuming anything further about the queries, the algorithm may not necessarily terminate.
\citet{DR14} do not discuss particular assumptions to consider for termination, so here we just consider one sufficient condition:
\begin{minipage}{\linewidth}
\begin{lstlisting}
abbrev has_lucky {sv_T : Type} (qs : sv_query sv_T) (T : ℤ) : Prop :=
  ∀ (τ : ℤ) (l : List sv_T), ∃ (K : ℤ), ∀ i, ∀ (K' : ℤ), K ≤ K' -> qs i l + K' ≥ τ + T
\end{lstlisting}
\end{minipage}
In other words, this requires that for each noise value $\tau$, there exists some value $K$ such that for each query in the stream, if we add at least $K$ to the value of the query on any database $l$, we get a value that exceeds the noised threshold.
This condition is true, for example, if the queries are all bounded below, i.e. there exists some minimum value $M$ such that $qs\; i\; l > M$ for all $i$ and $L$.
At a high level, this condition suffices to ensure termination because it means that there is some $\beta > 0$ such that for each query, with probability at least $\beta$, the noise added to the result of the query will cause it to exceed $\tau + T$.

This assumption is only used to prove almost-sure termination, and is not used in the rest of the privacy argument.
Therefore, an alternate termination assumption could be used without having to adjust the rest of the proof of privacy. \\

\paragraph{Returning Multiple Queries}
We have shown that \lstinline{sv1_aboveThresh}, the program corresponding to Algorithm 1 from \citet{DR14}, is $\epsilon$-DP.
However, this only returns the \emph{first} query that exceeds the threshold.
Based on this algorithm, \citet{DR14} then explain how to generalize the idea to compute the first $c$ queries which exceed the threshold in an algorithm that they call Algorithm 2.

\begin{figure}
\begin{lstlisting}[caption={The Sparse Vector Technique, implemented in terms of \lstinline{sv1_aboveThresh}.}, label=fig:sparsevector-sparse]
def snoc {T : Type*} (x : T) (L : List T) := L ++ [x]

def shift_qs (n : ℕ) (qs : sv_query sv_T) : sv_query sv_T := fun i => qs (i + n)

def privSparseAux {sv_T : Type} (qs' : sv_query sv_T) (HL' : has_lucky qs' T) (c : ℕ) : Mechanism sv_T (List ℕ) :=
  match c with
  | 0 => privConst []
  | Nat.succ c' =>
      privPostProcess
        (privComposeAdaptive
          (sv1_aboveThresh_PMF qs' T HL' ε₁ ε₂)
          (fun n => privSparseAux (shift_qs n qs') (shift_qs_lucky T n qs' HL') c'))
      (Function.uncurry snoc)

def privSparse (c : ℕ) : Mechanism sv_T (List ℕ) :=
  privSparseAux T ε₁ ε₂ (shift_qs 0 qs) (shift_qs_lucky _ _ _ HL) c

\end{lstlisting}
\end{figure}
\Cref{fig:sparsevector-sparse} depicts our implementation of Algorithm 2.\footnote{In particular, we implement the equivalent algorithm described in the text of their Theorem 3.25.}
To obtain a sequence of the first $c$ queries which exceed $T$, Algorithm 2 executes \lstinline{sv1_aboveThresh} a total of $c$ times, each time incuring a privacy cost of $\epsilon$.
The idea is that, if the $n$th iteration of \lstinline{sv1_aboveThresh}, returned index $k$, then on the $n+1$ iteration, we call \lstinline{sv1_aboveThresh} after \emph{shifting} the the $qs$ query function, so that it starts from query $k+1$.

While the proof we have seen so far for \lstinline{sv1_aboveThresh} is a pure-DP proof, the DP proof for the full sparse vector mechanism can be done parametrically in terms of the bound on \lstinline{sv1_aboveThresh}.
Therefore, our privacy proof for \lstinline{privSparse} is stated for a generic DP mechanism and is paramaterized by a specification for \lstinline{sv1_aboveThresh_PMF}, as shown in \cref{fig:sparsevector-param}.
Demonstrating that \lstinline{privSparseAux} is $(c \cdot \epsilon)$-DP follows by induction on $c$ and application of our generic postprocessing and adaptive composition bounds.
In fact, the code of \lstinline{privSparse} and \lstinline{privSparseAux} is similar in form to \lstinline{privNoisedHistogram} and \lstinline{privNoisedHistogramAux} in \cref{fig:histogram}, and their proofs are nearly identical as well.

\begin{figure}
\begin{lstlisting}[caption={Specifying a privacy bound for \lstinline{sv1_aboveThresh_PMF}, and an abstract proof of the privacy for \lstinline{privSparse_DP}.}, label=fig:sparsevector-param]
variable (T : ℤ) (ε₁ ε₂ : ℕ+) {sv_T : Type}
variable [dps : DPSystem sv_T]
variable [dpn : DPNoise dps]
variable (qs : sv_query sv_T)
variable (Hqs : has_lucky qs T)
variable (HDP : ∀ N H, ∀ ε : NNReal, (ε = ε₁ / ε₂) ->
  dps.prop (sv1_aboveThresh_PMF (shift_qs N qs) T H ε₁ ε₂) ε)

lemma privSparseAux_DP (ε : NNReal) (c : ℕ) (Hε : ε = ε₁ / ε₂) :
    ∀ N : ℕ, ∀ H, dps.prop (privSparseAux T ε₁ ε₂ (shift_qs N qs) H c) (c * ε) := by
  induction c
  · intro _ _
    unfold privSparseAux
    simp
    apply dps.const_prop
    simp_all
  · rename_i c' IH
    intro N HL
    simp [privSparseAux]
    apply dps.postprocess_prop
    apply @DPSystem.adaptive_compose_prop _ _ _ _ _ _ _ _ ε (c' * ε) ((c' + 1) * ε)
    · apply HDP
      trivial
    · intro u
      let IH' := IH (u + N)
      rw [<- shift_qs_add] at IH'
      apply IH'
    · ring_nf

lemma privSparse_DP (ε : NNReal) (c : ℕ) (Hε : ε = ε₁ / ε₂) :
    dps.prop (privSparse T ε₁ ε₂ qs Hqs c) (c * ε) := by
  unfold privSparse
  apply privSparseAux_DP
  · apply HDP
  · trivial
\end{lstlisting}
\end{figure}

\subsection{Establishing zCDP}

Rather than proving zCDP of the sparse vector technique directly, we instead chose to mechanize proposition 1.4 from \cite{bun2016concentrated} which states that every $\epsilon$-DP mechanism is $(\epsilon^{2}/2)$-zCDP.

We refer the interested reader to our development, since a low-level summary of our proof would only replicate the argument given in the proof of proposition 3.3 and lemma B.1 of \cite{bun2016concentrated}.
The mechanization of our proof involves using the privacy loss random variable formulation of the Rényi divergence, and Jensen's inequality to reduce the problem to showing
\begin{equation}
  0 \le y < x \le 2 \Rightarrow \frac{\sinh(x) - \sinh(y)}{\sinh(x-y)} \le e^{xy/2}.
\end{equation}

Like in \cite{bun2016concentrated} we prove this inequality by comparing the derivatives of the left- and right-hand side and concluding with the mean value theorem.
Mathlib already includes mechanized implementations for the calculus of hyperbolic functions, so this proof required little additional work on our part.

\section{Parallel Composition}\label{section:parallelcomposition}

Our choice to use \lean{} typeclasses in our definition of abstract DP means that it is open for extension: one can construct a hierarchy of DP definitions on top of AbstractDP in order to model versions of DP with more features.
One such example is \emph{parallel composition}: a technique common to many definitions of DP wherein one can obtain a privacy bound on a program which applies different DP mechanisms to disjoint subsets of the input dataset.
To demonstrate how \lstinline{AbstractDP} can be extended to support new generic DP features, we will create an extension for generic parallel composition, and use it to implement a differentially private histogram like in \cref{sec:histogram} but with a tighter generic bound.

\begin{figure}
\begin{lstlisting}[caption={Generic parallel composition}, label=fig:parcomp]
def privParComp (m1 : Mechanism T U) (m2 : Mechanism T V) (f : T -> Bool) : Mechanism T (U × V) :=
  fun l => do
    let v1 <- m1 <| List.filter f l
    let v2 <- m2 <| List.filter ((! ·) ∘ f) l
    return (v1, v2)
\end{lstlisting}
\end{figure}

\begin{figure}
\begin{lstlisting}[caption={An extension to AbstractDP for parallel composition.}, label=fig:abstractParDP]
class AbstractParDP (T : Type) extends (AbstractDP T) where
  prop_par {m₁ : Mechanism T U} {m₂  : Mechanism T V} {γ₁ γ₂ γ : NNReal} :
    γ = max γ₁ γ₂ -> ∀f, prop m₁ γ₁ -> prop m₂ γ₂ -> prop (privParComp m₁ m₂  f) γ
\end{lstlisting}
\end{figure}

\Cref{fig:parcomp} depicts a generic $\slangM{}$ implementation for 2-way parallel composition.
The program uses a predicate \lstinline{f} to partition the database, and applies a different mechanism to each of the partite sets.
For several common definitions of DP, if \lstinline{m1} is $\epsilon_{1}$-DP and \lstinline{m2} is $\epsilon_{2}$-DP, their parallel composition using any partition function will be $\max(\epsilon_{1}, \epsilon_{2})$-DP.
\Cref{fig:abstractParDP} shows a typeclass extension to \lstinline{AbstractDP} which describes this generic upper bound.
Note that the typeclass does not need to redefine \lstinline{prop} or any of the other generic DP rules.

In fact, implementing \lstinline{AbstractParDP} for a definition of DP only involves proving \lstinline{prop_par}.
For pure DP, the proof follows from the fact that for neighbouring databases (that is, input lists that differ by the inclusion or exclusion of a single row) at least one component of the tuple does not change.
By simplifying the \lstinline{privParComp} we see that at each point $(u, v)$ in \lstinline{U × V}:
\begin{equation}
  \textrm{\lstinline{privParComp}}\; m_{1}\; m_{2}\; f\; (u, v) = m_{1}(u) \cdot m_{2} (v)
\end{equation}
so in particular, whichever factor is constant will cancel out in the quotient for \lstinline{PureDP}, leaving the expression for the pure DP of the other mechanism.
This establishes an \lstinline{AbstractParDP} instance for pure DP. \\

\paragraph{Abstract Parallel Histogram.}
Parallel composition can allow us to obtain more accurate results by reducing the amount of noise necessary for privacy.
A common application is constructing a differentially private histogram, like in \cref{sec:histogram}, but by using parallel composition to count and add noise the values in each bin, allowing us to reduce the amount of noise by a factor of \lstinline{nBins} and still obtain the same privacy bound.

\begin{figure}
\begin{lstlisting}[caption={Generic parallel histogram}, label=fig:parHistogram]
variable [dps : AbstractParDP T] [dpn : DPNoise dps.toAbstractDP]
variable (numBins : ℕ+) (B : Bins T numBins)
def privParNoisedBinCount (γ₁ γ₂ : ℕ+) (b : Fin numBins) : Mechanism T ℤ :=
  (dpn.noise (exactBinCount numBins B b) 1 γ₁ γ₂)
def privParNoisedHistogramAux (γ₁ γ₂ : ℕ+) (n : ℕ) (Hn : n < numBins) : Mechanism T (Histogram T numBins B) :=
  let privParNoisedHistogramAux_rec :=
    match n with
    | Nat.zero => privConst (emptyHistogram numBins B)
    | Nat.succ n' => privParNoisedHistogramAux γ₁ γ₂ n' (Nat.lt_of_succ_lt Hn)
  privPostProcess
    (privParComp
      (privParNoisedBinCount numBins B γ₁ γ₂ n)
      privParNoisedHistogramAux_rec
      (B.bin · = n))
    (fun z => setCount numBins B z.2 n z.1)
def privParNoisedHistogram (γ₁ γ₂ : ℕ+) : Mechanism T (Histogram T numBins B) :=
  privParNoisedHistogramAux numBins B γ₁ γ₂ (predBins numBins) (predBins_lt_numBins numBins)
\end{lstlisting}
\end{figure}

Such an implementation is shown in \cref{fig:parHistogram}.
The main differences between the parallel histogram \lstinline{privParNoisedHistogram} and the sequential implementation in \cref{sec:histogram} is the usage of \lstinline{privParComp} rather than \lstinline{privCompose}, the reduction in the amount of noise added to each bin, and the restriction that the \lstinline{dps} be an instance of \lstinline{AbstractParDP } rather than \lstinline{AbstractDP}.
Using an almost identical proof to \cref{sec:histogram} (applying \lstinline{prop_par} rather than \lstinline{adaptive_compose_prop}) we are able to prove that \lstinline{privParNoisedHistogram} is $(\gamma_{1} / \gamma_{2})$-ADP.
As a consequence, this implementation is $(\gamma_{1} / \gamma_{2})$ pure DP.
For definitions of DP which may not meet the parallel composition bound but still satisfy \lstinline{AbstractDP}, the noisier but more general construction presented in \cref{sec:histogram} still applies: our choice to support parallel composition after the fact does not restrict the generality of our framework.

\section{Extraction and Compilation Details}\label{section:moreperformance}

In this appendix, we go into the technical details of our evaluation.

As mentioned in \cref{section:extraction}, we have two methods for obtaining executable code from \slang{} programs.
The first option is compilation using the foreign function interface; the full source code of our external functions is included in \cref{lst:cppfull}.

\begin{figure}[t]
  \footnotesize
\begin{lstlisting}[caption={The full C++ implementation of the external functions. }, label=lst:cppfull]
#include <lean/lean.h>
#include <fcntl.h>
#include <unistd.h>
#include <random>
static int urandom = -1;
extern "C" lean_object * prob_UniformByte (lean_object * eta) {
    lean_dec(eta);
    unsigned char r;
    read(urandom, &r,1);
    return lean_box((size_t) r); }
extern "C" lean_object * prob_Pure(lean_object * a, lean_object * eta) {
    lean_dec(eta);
    return a; }
extern "C" lean_object * prob_Bind(lean_object * f, lean_object * g, lean_object * eta) {
    lean_dec(eta);
    lean_object * exf = lean_apply_1(f,lean_box(0));
    lean_object * pa = lean_apply_2(g,exf,lean_box(0));
    return pa; }
extern "C" lean_object * prob_While(lean_object * condition, lean_object * body, lean_object * init, lean_object * eta) {
    lean_dec(eta);
    lean_object * state = init;
    lean_inc(state);
    lean_inc(condition);
    uint8_t cond = lean_unbox(lean_apply_1(condition,state));
    while (cond) {
        lean_inc(body);
        state = lean_apply_2(body,state,lean_box(0));
        lean_inc(condition);
        lean_inc(state);
        cond = lean_unbox(lean_apply_1(condition,state)); }
    return state; }
extern "C" lean_object * my_run(lean_object * a) {
    if (urandom == -1) {
        urandom = open("/dev/urandom", O_RDONLY | O_CLOEXEC);
        if (urandom == -1) {
            lean_internal_panic("prob_UniformByte: /dev/urandom cannot be opened"); } }
    lean_object * comp = lean_apply_1(a,lean_box(0));
    lean_object * res = lean_io_result_mk_ok(comp);
    return res; }
extern "C" uint32_t dirty_io_get(lean_object * a) {
    lean_object * r1 = lean_apply_1(a,lean_box(0));
    lean_object * r2 = lean_io_result_get_value(r1);
    if (lean_is_scalar(r2)) {
        size_t r3 = lean_unbox(r2);
        return r3; }
    else {
        lean_internal_panic("dirty_io_get: value not scalar"); } }
\end{lstlisting}
\end{figure}

The first five functions in \cref{lst:cppfull} (i.e., excluding \lstinline{dirty_io_get}) suffice to mark \slang{} programs as \textit{computable}, and to run \slang{} computations inside the \lean{} \lstinline{IO} monad.
The last function is used to expose a Python API to the C++ code, which we will revisit at the end of this section.

\begin{figure}[t]
  \footnotesize
\begin{lstlisting}[caption={A representative sample of our extracted code.}, label=lst:pythonBernExpNeg]
def BernoulliExpNegSample(self, num, den):
    o: bool = False
    if (num) <= (den):
        d_0_X_: bool
        out0_: bool
        out0_ = (self).BernoulliExpNegSampleUnit(num, den)
        d_0_X_ = out0_
        o = d_0_X_
    elif True:
        d_1_gamf_: int
        d_1_gamf_ = _dafny.euclidian_division(num, den)
        d_2_B_: bool
        out1_: bool
        out1_ = (self).BernoulliExpNegSampleGenLoop(d_1_gamf_)
        d_2_B_ = out1_
        if (d_2_B_) == (True):
            d_3_X_: bool
            out2_: bool
            out2_ = (self).BernoulliExpNegSampleUnit(_dafny.euclidian_modulus(num, den), den)
            d_3_X_ = out2_
            o = d_3_X_
        elif True:
            o = False
    return o
\end{lstlisting}
\end{figure}

The other technique for executing \slang{} programs is via a translation to Python. \cref{lst:pythonBernExpNeg} depicts a representative example of our translated code.
In order to produce Python code, we use a Lean metaprogram to inspect the syntax tree of a \slangM{} term and translate a limited subset of \lean{} directly.
Concretely, our program emits Dafny \cite{Dafny} source code, and then we use Dafny's existing support for compiling to Python.
Unlike the FFI, the extraction approach does not support arbitrary \textit{computable} \lean{} terms, however, translation to Python source code is advantageous for certain deployment settings.
In both cases, we rely on some trusted computing base, either the FFI implementation or the Dafny extraction, and all the toolchains needed to compile their respective output.\footnote{We did, however, check the distributions of the generated samples using a Kolmogorov–Smirnov test. The test for the extracted Python code was written in Python, and the test for the FFI compiled code was written inside \lean{}.}

\begin{center}
\begin{figure}
  \centering
  \includegraphics[scale=0.4]{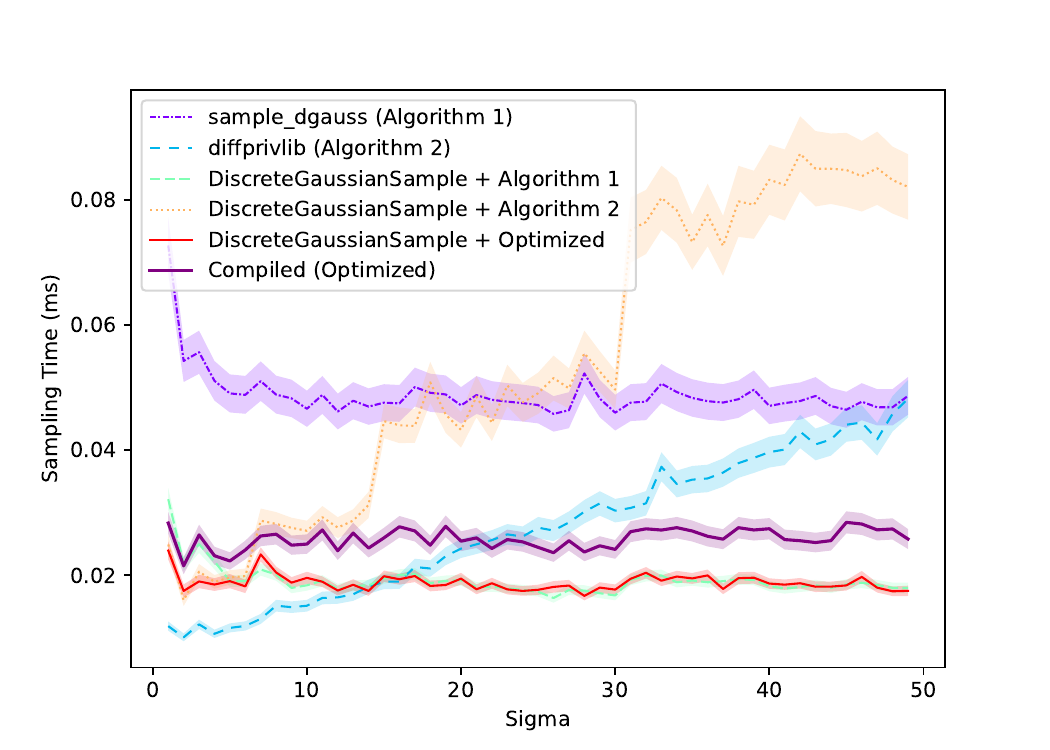}
  \caption{Runtime of \sampcert{}'s Gaussian sampler, with a C++ sampler included}\label{fig:samplingC}
\end{figure}
\end{center}

The benchmarks in \cref{fig:sampling} use the discrete Gaussian samplers extracted to Dafny and compiled to Python source code.
As alluded to previously, it is also possible to execute the compiled binary from inside Python using Python's foreign function interface.
\cref{fig:samplingC} depicts a benchmarking graph with calls to our C++ samplers through Python's FFI included.
Performance using the FFI samplers is competitive with the extracted samplers.

\begin{center}
\begin{figure}
  \centering
  \includegraphics[scale=0.4]{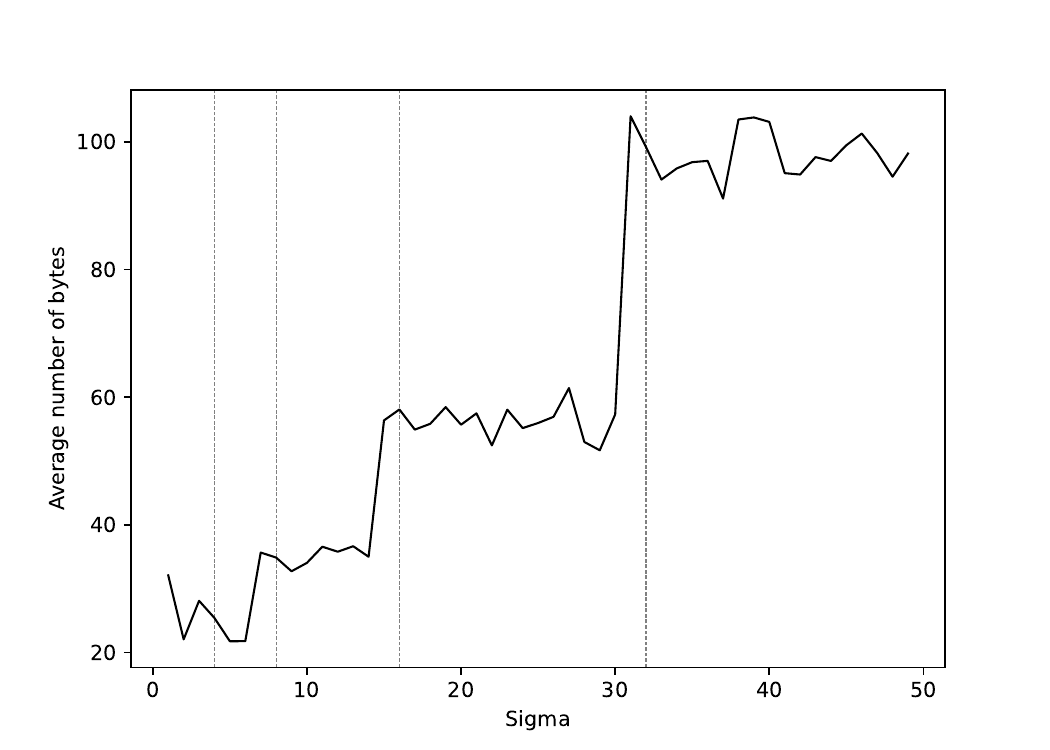}
  \caption{Number of random bytes read by the \sampcert{} implementation of Algorithm 2. Vertical lines in the graph are drawn at successive powers of 2. }\label{fig:samplingProfile}
\end{figure}
\end{center}

As already alluded to in \cref{section:performance}, our implementation of algorithm 2 has steps in its runtime, whereas the implementation in \lstinline{diffprivlib} does not.
We believe that the reason for this is due to differences in our uniform sampling algorithms: \lstinline{diffprivlib} uses uniform sampling based on randomly sampling floating-point numbers, whereas our exact uniform sampling algorithm consumes uniform random bytes proportional to the smallest number of bytes that is greater than its argument.
\Cref{fig:samplingProfile} is a graph of the number of bytes of entropy read by our implementation of Algorithm 2 while computing \cref{fig:sampling}, which we can see has similar spikes as the running time graph for our implementation of Algorithm 2.

\end{document}